\PassOptionsToPackage{unicode}{hyperref}
\PassOptionsToPackage{hyphens}{url}
\documentclass[
]{article}
\usepackage{xcolor}
\usepackage[margin=1in]{geometry}
\usepackage{amsmath,amssymb}
\usepackage{iftex}
\ifPDFTeX
  \usepackage[T1]{fontenc}
  \usepackage[utf8]{inputenc}
  \usepackage{textcomp} 
\else 
  \usepackage{unicode-math} 
  \defaultfontfeatures{Scale=MatchLowercase}
  \defaultfontfeatures[\rmfamily]{Ligatures=TeX,Scale=1}
\fi
\usepackage{lmodern}
\ifPDFTeX\else
\fi
\IfFileExists{upquote.sty}{\usepackage{upquote}}{}
\IfFileExists{microtype.sty}{
  \usepackage[]{microtype}
  \UseMicrotypeSet[protrusion]{basicmath} 
}{}
\makeatletter
\@ifundefined{KOMAClassName}{
  \IfFileExists{parskip.sty}{%
    \usepackage{parskip}
  }{
    \setlength{\parindent}{0pt}
    \setlength{\parskip}{6pt plus 2pt minus 1pt}}
}{
  \KOMAoptions{parskip=half}}
\makeatother
\usepackage{longtable,booktabs,array}
\usepackage{caption}
\usepackage{calc} 
\usepackage{etoolbox}
\makeatletter
\patchcmd\longtable{\par}{\if@noskipsec\mbox{}\fi\par}{}{}
\makeatother
\IfFileExists{footnotehyper.sty}{\usepackage{footnotehyper}}{\usepackage{footnote}}
\makesavenoteenv{longtable}
\usepackage{graphicx}
\makeatletter
\newsavebox\pandoc@box
\newcommand*\pandocbounded[1]{
  \sbox\pandoc@box{#1}%
  \Gscale@div\@tempa{\textheight}{\dimexpr\ht\pandoc@box+\dp\pandoc@box\relax}%
  \Gscale@div\@tempb{\linewidth}{\wd\pandoc@box}%
  \ifdim\@tempb\p@<\@tempa\p@\let\@tempa\@tempb\fi
  \ifdim\@tempa\p@<\p@\scalebox{\@tempa}{\usebox\pandoc@box}%
  \else\usebox{\pandoc@box}%
  \fi%
}
\def\fps@figure{htbp}
\makeatother
\providecommand{\tightlist}{%
  \setlength{\itemsep}{0pt}\setlength{\parskip}{0pt}}
\usepackage{graphicx}
\graphicspath{{figures/}{./}}
\usepackage{bookmark}
\IfFileExists{xurl.sty}{\usepackage{xurl}}{} 
\makeatletter
\@ifundefined{xmpquote}{}{}
\makeatother
\hypersetup{
  hidelinks,
  pdfcreator={LaTeX via pandoc}}

\author{}
\date{\vspace{-2.5em}}

\begin{document}

\section{Tournaments not inducible by five
voters}\label{tournaments-not-inducible-by-five-voters}

\emph{In loving memory of Adam Paton.}

\textbf{Leonid Chindelevitch\(^{a}\) and Ararat Harutyunyan\(^{b}\)}

\(^{a}\) MRC Centre for Global Infectious Disease Analysis, School of
Public Health, Imperial College London, 90 Wood Lane, London W12 0BZ,
United Kingdom. Email: \texttt{lchindel@ic.ac.uk}. ORCID:
0000-0002-6619-6013.

\(^{b}\) University of Paris-Dauphine, PSL University, CNRS UMR7243,
LAMSADE, Place du Maréchal de Lattre de Tassigny, 75775 Paris Cedex 16,
France. Email: \texttt{ararat.harutyunyan@lamsade.dauphine.fr}.

Corresponding author: Leonid Chindelevitch (\texttt{lchindel@ic.ac.uk}).

\subsection{Abstract}\label{abstract}

How many members must a committee have before its pairwise majorities
can express any collective preference? With enough voters every pattern
of pairwise majorities, or tournament, can arise, but a committee of
\(k\) voters produces only some; for odd \(k\), let \(N(k)\) be the
least number of alternatives on which some tournament arises from no
\(k\) voters. Only \(N(3) = 8\) is known; for five voters our companion
paper gave \(12 \le N(5) \le 38\) and the first explicit example of
moderate size, on 43 vertices. We show \(13 \le N(5) \le 22\). The upper
bound comes from nine tournaments on 22 vertices that five voters cannot
produce: eight one-vertex deletions of doubly regular tournaments on 23
vertices, including every deletion of one of them, and one arc reversal.
The Paley tournament on 23 vertices, which Bachmeier et al.~(2019) left
open after six weeks of SAT solving, is beyond five voters, and so is
every other doubly regular tournament on 23 vertices. The lower bound
comes from an analysis of all tournaments on 12 vertices. Both doubly
regular tournaments on 19 vertices, the Paley tournament and one other,
arise from five voters, but not with unit margin, that is, not with
every majority carried by three votes to two. The method is a search
that places one alternative at a time and exploits the tournament's
symmetries; four of the negative results are certified by
machine-checkable proofs, and all results are reproducible from a public
repository.

\textbf{Keywords:} majority voting; tournaments; McGarvey's theorem;
Paley tournaments; constraint search; SAT

\subsection{1. Introduction}\label{introduction}

Let \(A\) be a finite set of alternatives. A \emph{profile} is a finite
list of voters, each of whom ranks \(A\) by a linear order. When the
number of voters is odd, the \emph{majority tournament} of the profile
has, for each pair of alternatives \(u\) and \(v\), the arc \(u \to v\)
if more than half of the voters rank \(u\) above \(v\), and the arc
\(v \to u\) otherwise. McGarvey (1953) proved that every tournament on
\(A\), that is, every choice of one arc for each pair, is the majority
tournament of some profile. So when the number of voters is
unrestricted, majority rule imposes no structure on collective
preference: every pattern of pairwise majorities occurs, including
patterns with cycles. McGarvey's construction uses two voters for each
pair of alternatives, so the number of voters it needs grows with the
number of alternatives, and how many voters a pattern needs is the
question this paper is about.

Once the number of voters is fixed, majority rule does impose structure.
A panel of \(k\) voters admits only finitely many profiles, so it can
produce only finitely many patterns of pairwise majority, and once the
number of alternatives is large enough most patterns are beyond it. A
pattern that some \(k\)-voter panel does produce is called
\emph{\(k\)-inducible}. The question is then one about small
deliberative bodies: on an agenda of \(n\) alternatives, is a committee
of \(k\) members expressive enough to realize every collective
preference one could write down, or are there patterns of majority
preference that \(k\) members can never produce, however they rank the
alternatives?

For three voters the answer is known. Every pattern on at most seven
alternatives is realizable, and on eight alternatives one is not, by the
work of Shepardson and Tovey (2009) and of Eggermont, Hurkens and
Woeginger (2013). We write \(N(k)\) for the least number of alternatives
at which some pattern escapes \(k\) voters, so that \(N(3) = 8\). For
five voters, the smallest committee size at which the question is open,
the answer was known only to lie between \(12\) and \(38\), and no
explicit pattern on fewer than \(43\) alternatives was known to escape
five voters.

This paper narrows that to \(13 \le N(5) \le 22\), and both ends are
constructive: we give explicit five-voter profiles for the lower bound
and explicit patterns for the upper bound. Five voters suffice on every
agenda of at most twelve alternatives: we verify this for all
\(\approx 9.0 \times 10^{8}\) patterns on eleven alternatives together
with all of their one-alternative extensions. And on twenty-two
alternatives there are patterns that five voters cannot produce. We find
eight of them, each obtained by removing one alternative from a
perfectly balanced pattern on twenty-three alternatives (a \emph{doubly
regular tournament}, in the terminology defined below). The most
symmetric of them consists of two copies of the Paley tournament on
eleven alternatives, joined by a ladder (Figure 1); the Paley
tournaments form a well-known infinite family, with one tournament for
every prime number of alternatives congruent to \(3\) modulo \(4\). The
first step was the Paley tournament \(P_{23}\) on twenty-three
alternatives, the instance that a general-purpose SAT solver failed to
decide in six weeks of computation (Bachmeier et al.~2019); our search
settles it in twenty-two hours on a laptop.

\begin{figure}[!htbp]
\centering
\includegraphics[width=0.62\linewidth]{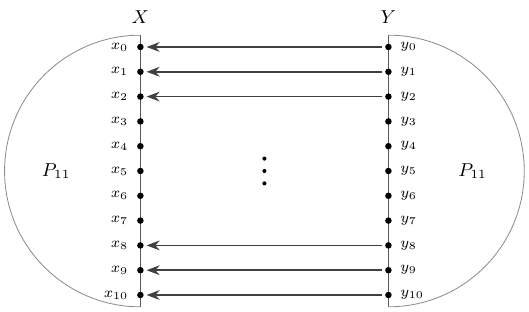}
\caption{The tournament $\Delta_1 - v_{10}$, which is not $5$-inducible. Its $22$ vertices split into $X = \{x_0, \dots, x_{10}\}$
and $Y = \{y_0, \dots, y_{10}\}$, with indices taken modulo $11$, and each half induces a copy of
$P_{11}$: $x_i \to x_j$ exactly when $j - i$ is a non-zero square modulo $11$, that is, lies in
$\mathrm{QR} = \{1, 3, 4, 5, 9\}$, and $y_i \to y_j$ exactly when $j - i$ is a non-square.
Between the halves, $x_i \to y_j$ when $j - i$ is a non-square, and $y_j \to x_i$ otherwise, that is,
when $j - i$ lies in $\mathrm{QR} \cup \{0\}$. In particular every rung $y_i \to x_i$ of the ladder
points from $Y$ to $X$; only six rungs are drawn. Each $x_i$ therefore has out-degree $10$
and each $y_j$ has out-degree $11$. Adding $1$ to every index, and multiplying every index
by an element of $\mathrm{QR}$, are automorphisms; together they generate the full automorphism group,
of order $55$.}
\end{figure}

The first pattern we found that five voters cannot produce, \(P_{23}\),
is as balanced as a collective preference on twenty-three alternatives
can be: every alternative defeats exactly eleven others, and every pair
of alternatives jointly defeats exactly the same number. It is the least
decisive pattern available, in the sense that no alternative and no pair
is distinguished from any other. What these obstructions share is
balance, not symmetry: \(P_{23}\) is also highly symmetric, but the
other \(36\) doubly regular tournaments on twenty-three alternatives are
just as balanced, \(32\) of them have no symmetry at all, and all of
them escape five voters too (Section 3.7).

Two further findings concern the margins by which majorities are
carried. A five-member panel can carry a pairwise majority by three
votes to two, by four to one, or unanimously, and it is natural to ask
whether the bare three-to-two margins suffice. They do not: \(P_{19}\)
is realizable by five voters, but every profile realizing it must carry
some majority by more than the bare minimum (Section 3.4). And the
failures are fragile in a precise sense: reversing a single pairwise
majority in \(P_{19}\) or in \(P_{23}\) restores realizability (Sections
3.3 and 3.6).

We use the definitions of our companion paper (Chindelevitch and
Harutyunyan 2026), in the notation below. Let \(T\) be a
\emph{tournament} on the vertex set \(V\), \(|V| = n\): a digraph in
which every unordered pair \(\{i,j\}\) carries exactly one of the arcs
\(i \to j\) or \(j \to i\). A \emph{\(k\)-profile} consists of \(k\)
linear orders \(\pi_1, \dots, \pi_k\) on \(V\). The \emph{majority
tournament} of such a profile puts an arc from \(i\) to \(j\) whenever
\(|\{ v : i \prec_{\pi_v} j \}| > k/2\). For odd \(k\) this is
well-defined on every pair, and the result is a tournament. We say \(T\)
is \emph{\(k\)-inducible} if some \(k\)-profile has \(T\) as its
majority tournament, and we call such a profile a \emph{witness} for
\(T\). The \emph{converse} of \(T\) is the tournament obtained from
\(T\) by reversing the direction of every arc.

We call the number of voters who rank \(i\) above \(j\) the
\emph{support} of the arc \(e: i \to j\), and denote it \(c(e)\). If
\(c(e) > k/2\), the \emph{margin} of \(e\) is \(2c(e) - k\). In a
witness every arc's support exceeds \(k/2\), so for \(k = 5\) it is
\(3\), \(4\) or \(5\). A witness has \emph{margin at most \(M\)} if
every arc has margin at most \(M\). For \(k = 5\) the possible margins
are \(1\), which we call \emph{unit margin}, \(3\) (every support in
\(\{3,4\}\)), and \(5\) (no restriction, i.e.~plain majority). A
unit-margin witness is in particular a majority witness. In our
companion paper we conjectured that the margin-based hierarchy is
strict, which for \(k = 5\) asserts two separate things: that some
tournament is \(5\)-inducible with margin at most \(3\) but not with
unit margin, and that some tournament is \(5\)-inducible but not with
margin at most \(3\). We settle the first here since \(P_{19}\) is such
a tournament (Section 3.4), while the second remains open (Section 4.2).

Reversing every voter's order reverses every arc of the majority
tournament and changes no support. So a profile realizes \(T\) exactly
when its reversal realizes the converse \(\bar{T}\), at the same
margins, and \(T\) and \(\bar{T}\) are \(k\)-inducible for the same
\(k\) and at the same margins. Every question in this paper is therefore
unchanged by taking the converse. Accordingly, we count a family
enumerated in full, such as all tournaments of a given order or all
vertex-transitive ones, up to isomorphism, and a family that our
searches generated up to isomorphism and converse.

One further quantity recurs below. Informally, the \emph{predictability}
\(\alpha^{*}(T)\) is the largest fraction \(\alpha > 1/2\) for which
there is some profile, on any number \(t\) of voters, such that every
arc of \(T\) is supported by at least \(\alpha t\) of them. Since every
arc of a \(k\)-voter witness is supported by more than \(k/2\) of the
voters, \(k\)-inducibility forces \(\alpha^{*}(T) \ge (k+1)/2k\), which
is \(2/3\) at \(k = 3\) and \(3/5\) at \(k = 5\). The implication is
only one-directional: \(\alpha^{*}\) can meet the threshold and the
tournament still fail to be inducible. Shepardson and Tovey (2009)
introduced the notion. Sections 5 and 7 of our companion paper compute
\(\alpha^{*}\) and gives, at both \(k = 3\) and \(k = 5\), tournaments
that meet the threshold but are not \(k\)-inducible.

Counting arguments show that almost all tournaments fail to be
\(k\)-inducible for any fixed \(k\), yet the \emph{explicit} examples
they yield are astronomically large. This paper closes much of the gap
for \(k = 5\). Let \(N(k)\) be the least order of a tournament that is
not \(k\)-inducible. The answer is only known exactly for \(k = 3\). For
\(k = 5\) the state of the art before this work was
\[12 \le N(5) \le 38,\] both ends of which are due to our companion
paper: the lower bound came from an exhaustive census of all
\(\approx 9.0 \times 10^{8}\) tournaments on \(11\) vertices, all of
which are \(5\)-inducible, improving on the bound \(N(5) \ge 11\) of
Bachmeier et al.~(2019).

The upper bounds in the literature are of two different kinds and they
are not always comparable. Some are non-constructive, and obtained by
counting: one compares the number of tournaments of a given order with
the number that \(k\) linear orders can produce, and when the first
exceeds the second, some tournament of that order must be missed.
Eggermont et al.~(2013) and Bachmeier et al.~(2019) argued this way to
reach \(41\), and our companion paper sharpened the comparison by
restricting it to regular and near-regular tournaments, reaching \(38\).
Other bounds are constructive, and exhibit an explicit tournament. The
previous record for \(k = 5\) was \(P_{43}\), from our companion paper,
the first explicit example of moderate order. In this paper, our search
shows that \(P_{23}\) is not \(5\)-inducible, which improves both the
best non-constructive (counting) and the best constructive (explicit)
upper bounds to \(N(5) \le 23\), and then that eight tournaments on
\(22\) vertices are not \(5\)-inducible either, so that \(N(5) \le 22\).

\subsubsection{1.1 Special properties of Paley
tournaments}\label{special-properties-of-paley-tournaments}

The Paley tournament \(P_q\), for a prime \(q \equiv 3 \pmod 4\), has
vertex set \(\mathbb{F}_q\) and an arc from \(i\) to \(j\) whenever
\(j - i\) is a nonzero square, or \emph{quadratic residue}; we write
\(\mathrm{QR}\) for the set of these, so that \(i \to j\) exactly when
\(j - i \in
\mathrm{QR}\). As everywhere in this paper, vertices are numbered from
\(1\), vertex \(i + 1\) being the field element \(i\). It is
\emph{doubly regular}: every vertex has out-degree \((q-1)/2\) and every
ordered pair of vertices has exactly \((q-3)/4\) common out-neighbours.
Its automorphism group, written \(\mathrm{Aut}(P_q)\) and formally
defined in Section 2, has order \(q(q-1)/2\). This makes \(P_q\)
\emph{arc-transitive}, meaning that for every pair of arcs \(e, e'\) of
\(P_q\) there is an automorphism of \(P_q\) that takes \(e\) into
\(e'\). It is also \emph{self-converse}, meaning that it is isomorphic
to its own converse; the relevant mapping is simply multiplication by
\(-1\), which is easily checked to not be an automorphism when
\(q \equiv 3 \pmod 4\).

\subsubsection{1.2 Why Paley(23) is hard to classify and Paley(43) was
easy}\label{why-paley23-is-hard-to-classify-and-paley43-was-easy}

Paley tournaments are natural candidates for non-inducibility because
they are far from being transitive, and because their symmetry makes an
otherwise hopeless search merely very hard. Bachmeier et al.~(2019) call
them \emph{quadratic residue tournaments} and write \(Q_q\); their
orientation is the converse of ours, which changes no result here
(Section 1); in particular a tournament and its converse have the same
\emph{majority dimension}, the least \(k\) for which a tournament is
\(k\)-inducible. We use the name Paley and the notation \(P_q\)
throughout.

The SAT study of Bachmeier et al.~(2019) establishes
\[\dim(Q_3) = \dim(Q_7) = 3, \qquad \dim(Q_{11}) = \dim(Q_{19}) = 5,\]
where \(\dim\) denotes majority dimension, and then stops exactly where
the problem becomes hard:

\begin{quote}
``Unfortunately, we were not able to check whether the majority
dimension of \(Q_{23}\) is equal to 5 or larger as the SAT solver did
not terminate within a total of six weeks.''
\end{quote}

By contrast, our previous proof that \(P_{43}\) is not \(5\)-inducible,
in our companion paper, never searched the profile space in general. It
rests on a quantity we call the \emph{slack},
\[\mathrm{slack}_5(T) = 5\,\mathrm{MAS}(T) - 3C, \qquad C = \binom{n}{2},\]
where \(\mathrm{MAS}\) is the maximum number of arcs of \(T\) that a
single linear order can agree with. Summing the majority condition over
all arcs bounds how far the five voters can \emph{collectively} fall
short of that maximum. A small slack restricts the voters to be close to
the best orders, and at \(q = 43\), where \(\mathrm{slack}_5 = 6\), that
restriction allowed us to fully enumerate about \(1.8\) million symmetry
classes of candidate orders.

While the same argument is valid at every \(q\), the slack is \emph{not}
a monotone function of \(q\) (table from our companion paper):

\begin{longtable}[]{@{}lllllll@{}}
\caption{The slack of the Paley tournaments, from our companion
paper.}\tabularnewline*
\toprule\noalign{}
\(q\) & 7 & 11 & 19 & 23 & 31 & 43 \\*
\midrule\noalign{}
\endfirsthead
\toprule\noalign{}
\(q\) & 7 & 11 & 19 & 23 & 31 & 43 \\*
\midrule\noalign{}
\endhead
\bottomrule\noalign{}
\endlastfoot
\(\mathrm{slack}_5\) & 7 & 10 & 22 & \textbf{46} & 30 & \textbf{6} \\*
\end{longtable}

Since \(P_{43}\) attains a locally minimal slack, while \(P_{23}\)
attains a local maximum, the method of our companion paper is unusable
at \(q = 23\). For any \(q\),
\[\mathrm{slack}_5(P_q) \;=\; 5C\left(\alpha^{*}(P_q) - 3/5\right),\]
where \(\alpha^{*}\) is the predictability; equality holds because
\(P_q\) is arc-transitive (see our companion paper). In particular, the
necessary condition \(\alpha^{*} \ge 3/5\) is far from binding, so no
argument resting on the predictability can certify infeasibility.

\textbf{Predictability separates \(k = 5\) from \(k = 3\).} At \(k = 3\)
predictability decides exactly at \(N(3)\): the \(96\) non-3-inducible
tournaments on eight vertices have \(\alpha^{*} = 13/20\), which is
\emph{below} \(2/3\), so the relaxation detects them, while all other
tournaments on eight vertices or fewer with at least one cycle have
\(\alpha^{*} = 2/3\) (Shepardson and Tovey 2009), and \(N(3) = 8\)
follows. At \(k = 5\) it does not: the table above has
\(\alpha^{*} > 3/5\) at every \(q\), yet \(P_{23}\), \(P_{31}\) and
\(P_{43}\) are all non-inducible, with \(P_{23}\) being the farthest
\emph{above} the threshold among the three.

\subsubsection{1.3 Contributions}\label{contributions}

\begin{enumerate}
\def\labelenumi{\arabic{enumi}.}
\item
  \textbf{There are non-\(5\)-inducible tournaments on \(22\) vertices,
  so \(N(5) \le 22\)} (Section 3.2). We write \(\Delta_i\) for the
  \(i\)-th of the \(37\) doubly regular tournaments on \(23\) vertices
  in the catalogue of McKay (2026), with vertices
  \(v_1, \dots, v_{23}\), and \(\Delta_i - v_j\) for \(\Delta_i\) with
  vertex \(v_j\) deleted. Eight such tournaments on \(22\) vertices,
  pairwise non-isomorphic and none the converse of another, are not
  \(5\)-inducible, and all eight are vertex-critical. Three of them are
  the three one-vertex deletions of \(\Delta_1\) up to isomorphism. One
  of these, \(\Delta_1 - v_{10}\), is two copies of \(P_{11}\) joined by
  a ladder, and a symmetry reduction using the automorphisms that map
  the base onto itself (Corollary 2.5) makes its refutation cheap;
  reversing one of its arcs gives a ninth obstruction (Section 3.3).
  \(\Delta_{11} - v_{20}\) also consists of two copies of \(P_{11}\),
  joined differently. The refutations of \(\Delta_1 - v_{10}\) and
  \(\Delta_6 - v_{5}\) are formally certified (Appendix B).
\item
  \textbf{\(P_{23}\) is not \(5\)-inducible} (Section 3.1). This settles
  the instance on which the SAT solver of Bachmeier et al.~(2019) did
  not terminate within a cumulative six weeks, and was the first step
  from \(38\) down to \(23\). We also decide \(P_{31}\), which does not
  improve the bound; the search settles it in less time than \(P_{23}\).
\item
  \textbf{\(P_{19}\) is \(5\)-inducible, but not \(5\)-inducible with
  unit margin} (Section 3.4). This is an example suggesting that
  \(k = 5\) may qualitatively differ from \(k = 3\). For \(k = 3\), the
  first \(k\)-inducible tournaments that are non-\(k\)-inducible with
  unit margin appear after \(N(k)\), order 9 vs order 8 (see our
  companion paper). For \(k = 5\) they appear no later than order
  \(19\), so the order is reversed unless \(N(5) \le 19\), which the
  search of Appendix G argues against.
\item
  \textbf{\(P_{19}\) and \(P_{23}\) are arc-critical} (Sections 3.3 and
  3.6). Reversing an arc in \(P_{19}\) restores \(5\)-inducibility at
  unit margin, and reversing an arc in \(P_{23}\) restores unrestricted
  \(5\)-inducibility.
\item
  \textbf{\(P_{31}\) and \(P_{43}\) are not vertex-critical} (Section
  3.5). Deleting a vertex from \(P_{31}\) or \(P_{43}\) leaves a
  tournament that is still not \(5\)-inducible.
\item
  \textbf{Four formally certified negative results} (Appendix B).
  \(P_{19}\) is not \(5\)-inducible at unit margin, and \(P_{23}\),
  \(\Delta_1 - v_{10}\) and \(\Delta_6 - v_{5}\) are not
  \(5\)-inducible. The search is split into independent subproblems
  (\emph{cubes}); each is refuted by the SAT solver CaDiCaL, which emits
  its refutation as a machine-checkable \emph{LRAT} proof; and every one
  of those proofs is then rechecked by a separate program that shares no
  code with the solver.
\item
  \textbf{Every tournament on \(12\) vertices is \(5\)-inducible, even
  at unit margin} (Appendix C). The analysis settles all \(2{,}048\)
  one-vertex extensions of each of the \(\approx 9.0 \times 10^{8}\)
  tournaments on \(11\) vertices, which is complete because every
  \(12\)-vertex tournament is such an extension of each of its twelve
  one-vertex deletions.
\end{enumerate}

\subsubsection{1.4 Related work}\label{related-work}

This paper builds on two others. Our companion paper (Chindelevitch and
Harutyunyan 2026) is the prior art for the \emph{bound}: it produced the
first explicit non-\(5\)-inducible tournament of modest size,
\(P_{43}\), and its two bounds are what we measure against:
\(N(5) \le 38\) by counting and \(N(5) \le 43\) explicitly. Bachmeier et
al.~(2019) is the prior art for the \emph{method}: it brought SAT to
bear on \(k\)-majority digraphs, characterized the \(3\)-majority case
building on Dushnik and Miller (1941), and established hardness results
for several voting rules with a constant number of voters.

The value of \(N(3)\) is settled. Shepardson and Tovey (2009) introduced
the predictability \(\alpha^{*}(T)\) defined above. A \(3\)-voter
profile is in particular a \(2/3\)-supermajority representation, so
\(\alpha^{*}(T) < 2/3\) rules out \(3\)-inducibility; their \(8\)-vertex
tournament has \(\alpha^{*} = 13/20\), which gives \(N(3) \le 8\). They
also prove that every tournament on at most \(7\) vertices has
\(\alpha^{*} \in \{2/3, 1\}\), and Eggermont, Hurkens and Woeginger
(2013) then show by direct computation using an ILP formulation that
every tournament on \(7\) vertices is \(3\)-inducible, and every
tournament on \(8\) or \(9\) vertices is \(5\)-inducible, yielding
\(N(3) = 8\) and \(N(5) \ge 10\).

Our certified computation (Appendix B) is standard. It splits the
problem into independent subproblems, using the \emph{cube-and-conquer}
paradigm (Heule et al.~2012): a SAT instance is split by fixing a few
variables at a time, giving many partial assignments called
\emph{cubes}, and each cube is handed to the solver as a problem of its
own. We adopt the standard of evidence set by the Boolean Pythagorean
triples proof (Heule et al.~2016), later re-checked with a formally
verified checker (Cruz-Filipe et al.~2019), in which every subproblem
emits a machine-checkable refutation that an independent checker
validates. Our refutations are in the LRAT format (Cruz-Filipe et
al.~2017) (Appendix B.1), a format in which the solver records each
clause it derives so that an independent checker can replay and confirm
the entire derivation.

This paper grew out of our companion paper, whose constructive bound
\(N(5) \le 43\) came from a structural argument, and whose counting
bound \(N(5) \le 38\) names no explicit tournament. The question of what
a decision procedure would reach at smaller orders led to the search of
Section 2.2.

The rest of this paper is organized as follows. Section 2 sets out the
notation, the symmetry reductions and the search in outline. Section 3
presents the results: \(P_{23}\) and \(P_{31}\) (Section 3.1), the eight
\(22\)-vertex obstructions that give \(N(5) \le 22\) (Section 3.2),
vertex- and arc-criticality (Sections 3.3 to 3.6), and other structured
families (Section 3.7). Section 4 discusses where the method may
transfer and what remains open, and Section 5 describes the
reproducibility material. Appendix A describes the search engine and the
cost of every computation, Appendix B the formal certificates, Appendix
C the analysis of order \(12\), Appendix D the exact enumeration counts,
Appendix E the construction for locally transitive tournaments, Appendix
F criticality at \(21\) vertices, Appendix G the search below \(22\)
vertices, and Appendix H the adjacency matrices of the \(22\)-vertex
obstructions.

\subsection{2. Preliminaries}\label{preliminaries}

Throughout, \(T\) is a tournament on \(V = \{1, \dots, n\}\) and
\(k = 5\) unless stated otherwise. We write \(N^{+}(v)\) and
\(N^{-}(v)\) for the out- and in-neighbourhoods of \(v\). A tournament
is \emph{regular} if \(|N^{+}(v)| = |N^{-}(v)|\) for each \(v \in V\);
this can only happen for \(n\) odd. Following our companion paper,
\(D_n\), \(R_n\) and \(S_n\) denote the numbers of isomorphism classes
of, respectively, all tournaments, regular tournaments and self-converse
tournaments on \(n\) vertices. The text gives them to two significant
figures and Appendix D tabulates their exact values. For any
\(S \subseteq V\), the sub-tournament of \(T\) induced by \(S\) is the
tournament obtained by restricting \(T\) to the vertices in \(S\) and
the arcs between those vertices.

We call a tournament \(T\) \emph{vertex-critical} for \(k\)-inducibility
at margin \(M\) if \(T\) is not \(k\)-inducible at margin \(M\), but
\(T - v\) is \(k\)-inducible at margin \(M\) for every vertex \(v\).
Similarly, we call \(T\) \emph{arc-critical} for such a property if
\(T\) has it, but any \(T'\) obtained from \(T\) by reversing an arc
does not.

\subsubsection{2.1 Symmetry reductions}\label{symmetry-reductions}

An \emph{automorphism} of a tournament \(T\) is a permutation of its
vertices carrying every arc to an arc; these form a group
\(\mathrm{Aut}(T)\) under composition. The identity is always an
automorphism; any other automorphism is \emph{non-trivial}, and \(T\) is
\emph{rigid} when it has none. For a permutation group \(\Gamma\) on a
set \(X\), the \emph{orbit} of \(x\) is
\(\Gamma x = \{\gamma(x) : \gamma \in \Gamma\}\), and the orbits
partition \(X\). The group is \emph{transitive} on \(X\) when there is a
single orbit, that is, when for any \(x, x' \in X\) there is a
\(g \in \Gamma\) with \(g(x) = x'\). We use orbits of
\(\mathrm{Aut}(T)\) on the vertices, on the arcs, and on the
\emph{ordered pairs} of distinct vertices, where \(\gamma\) sends
\((u,v)\) to \((\gamma(u), \gamma(v))\). An ordered pair is either an
arc or the reverse of one, so that third action has at least two orbits.

Two symmetries act on witnesses and both are exploited by every method
here: the \(k!\) relabellings of the voters, which form the symmetric
group \(S_k\), and \(\mathrm{Aut}(T)\) acting on the vertices. A
permutation \(\sigma\) of the voters acts on a profile
\(\pi = (\pi_1, \dots, \pi_k)\) by reordering it, and a permutation
\(h\) of \(V\) acts on a linear order by relabelling, so that \(h\pi_j\)
ranks \(h(u)\) above \(h(w)\) exactly when \(\pi_j\) ranks \(u\) above
\(w\), and on a profile voter by voter. The two actions commute, so the
group \(S_k \times \mathrm{Aut}(T)\) acts on profiles. It maps witnesses
to witnesses: permuting voters does not change any margin, and the
margin of \(h(u) \to h(w)\) after relabelling by an automorphism \(h\)
equals that of \(u \to w\) before. Every reduction we use is a special
case of one principle.

\textbf{Theorem 2.1 (orbit principle).} \emph{Let \(G\) be a subgroup of
\(S_k \times \mathrm{Aut}(T)\) acting on a set \(\Omega\), and let
\(\varphi\) map profiles to \(\Omega\) with
\(\varphi(g\pi) = g\,\varphi(\pi)\) for every \(g \in G\). Let \(R\) be
a subset of \(\Omega\). Then if \(R\) contains an element of every
\(G\)-orbit and \(T\) is \(k\)-inducible, \(T\) has a witness \(\pi\)
with \(\varphi(\pi) \in R\).}

\emph{Proof.} Let \(w\) be a witness and choose \(g \in G\) with
\(g\,\varphi(w) \in R\). Then \(gw\) is a witness and
\(\varphi(gw) = g\,\varphi(w) \in R\). \(\blacksquare\)

We never apply Theorem 2.1 directly. What the methods use are its
corollaries below; the theorem gives them a common proof and tells us
when two of them can be combined.

The corollaries below differ only in the group \(G\) and in the object
\(\varphi(\pi)\) that a profile is reduced to (Table 1). The group is
often a subgroup because the object dictates it: \(\mathrm{Aut}(T)\)
acts on the base states of \(B\) only through the automorphisms that map
\(B\) onto itself.

\begin{longtable}[]{@{}
  >{\raggedright\arraybackslash}p{(\linewidth - 4\tabcolsep) * \real{0.2273}}
  >{\raggedright\arraybackslash}p{(\linewidth - 4\tabcolsep) * \real{0.3182}}
  >{\raggedright\arraybackslash}p{(\linewidth - 4\tabcolsep) * \real{0.4545}}@{}}
\caption{The symmetry reductions of this section as instances of Theorem
2.1. Here \(B\) is the base and \(H\) is a subgroup of
\(\mathrm{Aut}(T)\) mapping a subset \(C \subseteq B\) onto
itself.}\tabularnewline*
\toprule\noalign{}
\begin{minipage}[b]{\linewidth}\raggedright
result
\end{minipage} & \begin{minipage}[b]{\linewidth}\raggedright
group \(G\)
\end{minipage} & \begin{minipage}[b]{\linewidth}\raggedright
object \(\varphi(\pi)\)
\end{minipage} \\*
\midrule\noalign{}
\endfirsthead
\toprule\noalign{}
\begin{minipage}[b]{\linewidth}\raggedright
result
\end{minipage} & \begin{minipage}[b]{\linewidth}\raggedright
group \(G\)
\end{minipage} & \begin{minipage}[b]{\linewidth}\raggedright
object \(\varphi(\pi)\)
\end{minipage} \\*
\midrule\noalign{}
\endhead
\bottomrule\noalign{}
\endlastfoot
Corollary 2.2 & \(\mathrm{Aut}(T)\) & the top two vertices of
\(\pi_1\) \\*
Corollary 2.3 & \(\mathrm{Aut}(T)\) & the top vertex of \(\pi_1\) \\*
Corollary 2.4 & \(S_k\) & the profile \(\pi\) itself \\*
Corollary 2.5 & \(S_k \times H\), \(h(C) = C\) & the base state of
\(C \subseteq B\) that \(\pi\) extends \\*
\end{longtable}

As a witness can be transformed by an automorphism, we may in some cases
decide in advance what some voter's top two vertices are.

\textbf{Corollary 2.2 (orbit anchoring).} \emph{Let \(\mathrm{Aut}(T)\)
act on the ordered pairs of distinct vertices, and choose one
representative pair from each orbit, say
\((a_1, b_1), \dots, (a_m, b_m)\). If \(T\) is \(k\)-inducible, then
\(T\) has a witness in which, for at least one index \(i\), some voter
ranks \(a_i\) first and \(b_i\) second.}

\emph{Proof.} Theorem 2.1 with \(G = \mathrm{Aut}(T)\), \(\varphi(\pi)\)
the ordered pair of the top two vertices of \(\pi_1\), and \(R\) the
chosen pairs. \(\blacksquare\)

Thus we may search only for witnesses whose top pair is one of the \(m\)
chosen pairs. The corollary only guarantees that \emph{some} \(i\)
works, not any particular one.

For a Paley tournament \(m = 2\), and the representatives can be
exhibited without computing \(\mathrm{Aut}(P_q)\). For
\(a \in \mathrm{QR}\) and any \(t\), the map \(x \mapsto a(x-t)\)
multiplies every difference by \(a\), which cannot change whether a
difference lies in \(\mathrm{QR}\), so it is an automorphism; it sends
the ordered pair \((t,s)\) to \((0, a(s-t))\), and as \(a\) ranges over
\(\mathrm{QR}\), \(a(s-t)\) ranges over the coset
\(\mathrm{QR}\cdot(s-t)\). Since \(\mathbb{F}_q^{*}\) is the disjoint
union of \(\mathrm{QR}\) and its complement, each of them one such
coset, these maps alone leave only two orbits, separated by whether the
difference lies in \(\mathrm{QR}\). The orbits of the full group are
unions of these, so there are at most two, and since automorphisms carry
arcs to arcs, at least two. They are therefore exactly the arcs and the
non-arcs, and any arc together with any non-arc serves as
representatives. On \(P_7\), worked through in Figure A.1 of Appendix
A.3, \(\mathrm{QR} = \{1,2,4\}\), so the representatives are \((1,2)\),
whose difference \(1\) lies in \(\mathrm{QR}\), and \((1,4)\), whose
difference \(3\) does not; both lie inside the base \(B = \{1,2,4\}\)
used there, which is what lets the anchoring be checked from the base
state alone, and panel (c) rejects an insertion that would lift a vertex
above an anchored pair.

\textbf{Corollary 2.3 (vertex anchoring).} \emph{Let \(\mathrm{Aut}(T)\)
act on the vertices and choose one representative \(v_1, \dots, v_r\)
from each orbit. If \(T\) is \(k\)-inducible, then \(T\) has a witness
in which some voter ranks some \(v_i\) first.}

\emph{Proof.} Theorem 2.1 with \(G = \mathrm{Aut}(T)\), \(\varphi(\pi)\)
the top vertex of \(\pi_1\), and \(R\) the chosen vertices.
\(\blacksquare\)

In both corollaries a \emph{representative} is one of the chosen
vertices or ordered pairs of vertices, one from each orbit of
\(\mathrm{Aut}(T)\); it is part of \(T\), not a base state. Corollaries
2.2 and 2.3 apply only if the representatives meet every orbit, so we
compute \(\mathrm{Aut}(T)\) and its orbits for every tournament, with
\texttt{nauty} 2.8.6 (McKay and Piperno 2014), which also provides every
automorphism group, orbit set and canonical form reported in this paper.
We say that a witness meeting either restriction is \emph{anchored}. The
two are alternatives rather than a sequence, each without loss of
generality, and which is better depends on the tournament's number of
orbits on vertex pairs compared with that on vertices, as well as the
relative complexity of testing each restriction.

\textbf{Corollary 2.4 (voter lex-ordering).} \emph{If \(T\) is
\(k\)-inducible, then \(T\) has a witness whose \(k\) orders are
non-decreasing under any fixed total order on linear orders.}

\emph{Proof.} Theorem 2.1 with \(G = S_k\), \(\varphi\) the identity,
and \(R\) the profiles whose orders are non-decreasing; every
\(S_k\)-orbit contains one, obtained by sorting. \(\blacksquare\)

Two reductions compose when the second does not undo the first. After
restricting to witnesses with \(\varphi_1(\pi) \in R_1\), a second
application of Theorem 2.1 may use only elements of its group that
preserve this property. Anchoring followed by sorting is such a pair:
every permutation of the voters preserves the property Corollary 2.2
guarantees, ``\emph{some} voter ranks \(a_i\) first and \(b_i\)
second'', which depends on the multiset of orders and not on which voter
carries it. Had it been sharpened to ``voter \(1\) ranks \(a_i\) first
and \(b_i\) second'', the composition would fail. Both are therefore
available to every formulation in Appendix A.

Together they reduce the running time substantially, in two different
ways. Corollary 2.4 is absorbed into the definition of a base state: the
\(k\) orderings of \(B\) are required to be non-decreasing, dividing the
count by almost \(120\) (almost, because a tuple with two equal orders
is fixed by some of the \(120\) permutations).

Corollary 2.2 acts earlier still, \emph{invalidating} base states
outright, but only when \emph{all} the representative pairs lie inside
\(B\). Suppose they do, and that a base state extends to an anchored
witness, some voter of which ranks \(a_i\) first and \(b_i\) second
overall. As \(a_i, b_i \in B\), that voter ranks \(a_i\) first and
\(b_i\) second \emph{within} \(B\), which the base state alone reveals.
Contrapositively, a base state no voter of which tops \(B\) with a
chosen pair extends to no anchored witness, hence to no witness at all,
and may be discarded. If one representative has an endpoint outside
\(B\) this filter does not apply; a base state failing the test on the
representatives it sees may still extend to a witness anchored on the
one it cannot. We call a base state \emph{live} if it survives. On the
five-vertex base we use on the Paley tournaments \(2{,}591\) of the
\(8{,}031\) are live.

The search and the certification differ in \emph{when} the condition in
Corollary 2.2 is tested. Both test it at the start to determine the live
base states. Our algorithmic approach keeps testing it during the
search, since a base state fixes the orders only on \(B\) whereas the
condition constrains the full orders: a voter picking \((a_i, b_i)\) as
its top pair in \(B\) ceases to be anchored once a later vertex is
inserted above \(b_i\). Pruning is sound once every representative has
both endpoints placed, for then the test is monotone. An insertion can
destroy an anchor but never create one, the placed vertices' relative
order being already fixed, so a partial profile with no anchored voter
has no extension with one. The SAT certification instead tests the
condition once, at the root, where it decides which cubes reach the
solver.

Corollaries 2.2 and 2.3 restrict, without loss of generality, what a
witness looks like on all of \(V\), and so how the base interacts with
the vertices outside it. When a subgroup of automorphisms maps the base
onto itself, the same idea works inside the base: the subgroup
identifies base states with one another, and only one from each class
needs to be searched. Every witness \(\pi_1, \dots, \pi_k\) determines a
base state of \(B\), namely the multiset of the restricted orders
\(\pi_1|_B, \dots, \pi_k|_B\) (Appendix A.2 gives the full definition),
and we say that the witness \emph{extends} this base state. The
\emph{restriction} of a base state of \(B\) to a subset
\(C \subseteq B\) is the multiset of the restrictions of its \(k\)
orders to \(C\), which is a base state of \(C\).

\textbf{Corollary 2.5 (base-stabilizer quotient).} \emph{Let \(H\) be a
subgroup of \(\mathrm{Aut}(T)\) and \(C \subseteq B\) with \(h(C) = C\)
for every \(h \in H\) (possibly \(C = B\)), and choose one base state of
\(C\) from each \(H\)-orbit. Call a base state of \(B\) chosen if its
restriction to \(C\) is chosen.}

\emph{(a) \(H\) permutes the base states of \(C\), and if one of them
extends to a witness, so does every base state in its \(H\)-orbit.
Consequently, if \(T\) is \(k\)-inducible, some chosen base state of
\(B\) extends to a witness.}

\emph{(b) Suppose in addition that every element of \(H\) fixes each
representative vertex of Corollary 2.3. Then, if \(T\) is
\(k\)-inducible, some chosen base state extends to a witness anchored in
the sense of Corollary 2.3.}

\emph{(c) Suppose instead that every element of \(H\) fixes both \(a_i\)
and \(b_i\) for every representative pair \((a_i, b_i)\) of Corollary
2.2. Then, if \(T\) is \(k\)-inducible, some chosen base state extends
to a witness anchored in the sense of Corollary 2.2.}

\emph{Proof.} As \(h(C) = C\), relabelling by \(h\) maps an order of
\(C\) to an order of \(C\), and since \(h\) restricts to an automorphism
of \(T|_C\), it maps a base state of \(C\) to a base state of \(C\);
this is the action of \(H\), and \(S_k\) acts trivially, base states
being multisets. The base state of \(C\) that \(h\pi\) extends is \(h\)
applied to the one \(\pi\) extends, so Theorem 2.1 with
\(G = S_k \times H\) and \(\varphi(\pi)\) that base state gives a
witness extending a chosen base state \(Q_C\) of \(C\); applying
\(h^{-1}\) gives the orbit statement. The restrictions
\(\pi_1|_B, \dots, \pi_k|_B\) of that witness form a base state \(Q_B\)
of \(B\) that it extends, and the restriction of \(Q_B\) to \(C\) is
\(Q_C\), so \(Q_B\) is chosen. For (b), an element of \(H\) maps a voter
ranking \(v_i\) first to one ranking \(h(v_i) = v_i\) first, so it
preserves anchoring, and the composition rule above applies. For (c),
likewise, an element of \(H\) maps a voter ranking \(a_i\) first and
\(b_i\) second to one ranking \(h(a_i) = a_i\) first and
\(h(b_i) = b_i\) second. \(\blacksquare\)

With \(C = B\), the search therefore needs only one base state of \(B\)
from each \(H\)-orbit, which divides the work by up to \(|H|\), and
Corollary 2.4 still holds because base states remain multisets of
orders. The condition on anchoring cannot be dropped: if \(H\) moves
some representative, mapping an anchored witness into the chosen base
state can destroy its anchor, and only one of the two reductions can be
used.

Finding \(H\) is a question about \(\mathrm{Aut}(T)\). Automorphism
groups of tournaments have odd order, since an automorphism of order
\(2\) would swap the two endpoints of some arc. By Cauchy's theorem,
\(\mathrm{Aut}(T)\) contains an element of order \(p\) for every prime
\(p\) dividing its order, and an element \(h\) maps \(B\) onto itself
exactly when \(B\) is a union of cycles of \(h\). When \(|B| = 5\) and
\(h\) has order \(5\), a single \(5\)-cycle of \(h\) is such a base, and
the subgroup \(H = \{\mathrm{id}, h, h^2, h^3, h^4\}\) of its powers
divides the number of base states by up to \(5\). The base states fixed
by \(h\) are the only ones whose orbit is smaller than \(|H|\).

A subgroup that maps a base of the size we search with onto itself need
not exist, and this is where \(C \subsetneq B\) helps. An automorphism
of order \(3\) with no fixed point, for example, maps no set of \(5\)
vertices onto itself, but it maps each of its \(3\)-cycles onto itself.
With \(C\) one such \(3\)-cycle inside a base \(B\) of \(5\) vertices,
the search keeps only the base states of \(B\) above a chosen base state
of \(C\). As \(H\) need not map \(B\) onto itself, it does not act on
the base states of \(B\), and the fraction kept is not exactly
\(1/|H|\); we choose in each orbit the base state of \(C\) with the
fewest base states of \(B\) above it. Three of the unit-margin searches
of Appendix G used this form, keeping \(600\) of the \(2{,}200\) base
states of a five-vertex base inducing the regular tournament on five
vertices.

\textbf{Using the whole group.} Anchoring and the quotient use different
parts of \(\mathrm{Aut}(T)\), in the way the orbit--stabilizer theorem
(Dummit and Foote 2004, Section 4.1, Proposition 2) describes. Anchoring
(Corollaries 2.2 and 2.3) uses the orbits of the chosen representatives,
and the quotient (Corollary 2.5) uses a subgroup that fixes every
representative. We therefore take \(H = G_R\), the subgroup of
\(\mathrm{Aut}(T)\) that fixes each representative, which meets the
condition of Corollary 2.5(b) or (c) by construction. For a single
representative \(r\), let
\(\mathrm{Aut}(T) \cdot r = \{h(r) : h \in \mathrm{Aut}(T)\}\) be its
orbit and \(G_r = \{h \in \mathrm{Aut}(T) : h(r) = r\}\) its stabilizer;
then \(|\mathrm{Aut}(T)| = |\mathrm{Aut}(T) \cdot r| \cdot |G_r|\), so
between them the two reductions involve the whole group: its transitive
part through anchoring and the stabilizer through the quotient. With
several representatives, \(G_R\) is the intersection of their
stabilizers and can be smaller, so we choose the representatives to keep
it large, and then choose the base among the unions of cycles of
\(G_R\), or bases containing such a union (Corollary 2.5 with
\(C \subsetneq B\)). The two parts do not save work at the same rate.
The quotient divides the number of base states by up to \(|G_R|\),
whereas anchoring discards base states only when every representative
lies in the base, and otherwise prunes during the search by an amount
that depends on the tournament. Table 2 shows the split for the
tournaments of Section 3. For the Paley tournaments the stabilizer of a
representative pair is trivial, so the whole group goes to anchoring and
the saving is well below \(|\mathrm{Aut}(T)|\); for
\(\Delta_1 - v_{10}\) the group splits into orbits of \(11\) for
anchoring and a stabilizer of order \(5\) for the quotient.

\begin{table}[!htbp]
\centering
\caption{How the searches and certificates of Section 3 use the
automorphism group. Anchoring uses the orbits of the representatives and
the quotient their common stabilizer. In the \(\Delta_1 - v_{10}\)
search the representatives lie outside the base, so anchoring prunes
during the search and the count of base states kept reflects the
quotient alone.}
\begin{tabular}{@{}
  >{\raggedright\arraybackslash}p{(\linewidth - 8\tabcolsep) * \real{0.2680}}
  >{\raggedleft\arraybackslash}p{(\linewidth - 8\tabcolsep) * \real{0.0928}}
  >{\raggedright\arraybackslash}p{(\linewidth - 8\tabcolsep) * \real{0.2680}}
  >{\raggedright\arraybackslash}p{(\linewidth - 8\tabcolsep) * \real{0.1134}}
  >{\raggedleft\arraybackslash}p{(\linewidth - 8\tabcolsep) * \real{0.2577}}@{}}
\toprule\noalign{}
\begin{minipage}[b]{\linewidth}\raggedright
tournament
\end{minipage} & \begin{minipage}[b]{\linewidth}\raggedleft
\(\lvert \mathrm{Aut}(T) \rvert\)
\end{minipage} & \begin{minipage}[b]{\linewidth}\raggedright
anchoring
\end{minipage} & \begin{minipage}[b]{\linewidth}\raggedright
quotient
\end{minipage} & \begin{minipage}[b]{\linewidth}\raggedleft
base states kept
\end{minipage} \\
\midrule\noalign{}

\(P_{19}\), unit, certificate & \(171\) & ordered pairs & none &
\(22{,}876\) of \(142{,}251\) \\
\(P_{23}\), search & \(253\) & ordered pairs & none & \(2{,}591\) of
\(8{,}031\) \\
\(P_{23}\), certificate & \(253\) & ordered pairs & none & \(343{,}896\)
of \(3{,}414{,}729\) \\
\(\Delta_1 - v_{10}\), search & \(55\) & \(x_0\), \(y_0\) (orbits of
\(11\)) & order \(5\) & \(1{,}611\) of \(8{,}031\) \\
\(\Delta_1 - v_{1}\), \(\Delta_1 - v_{2}\) & \(5\) & none & order \(5\)
& \(1{,}611\) of \(8{,}031\) \\
\(\Delta_{11} - v_{2}\), \(- v_{5}\), \(- v_{20}\) & \(5\) & none &
order \(5\) & \(1{,}611\) of \(8{,}031\) \\
\(\Delta_6 - v_{5}\), \(\Delta_{28} - v_{5}\) & \(1\) & none & none &
\(8{,}031\) of \(8{,}031\) \\
\bottomrule\noalign{}
\end{tabular}
\end{table}

Figure 2 summarizes which of these reductions apply to a given
tournament and base, and how they combine.

\begin{figure}[!htbp]
\centering
\includegraphics[width=0.78\linewidth]{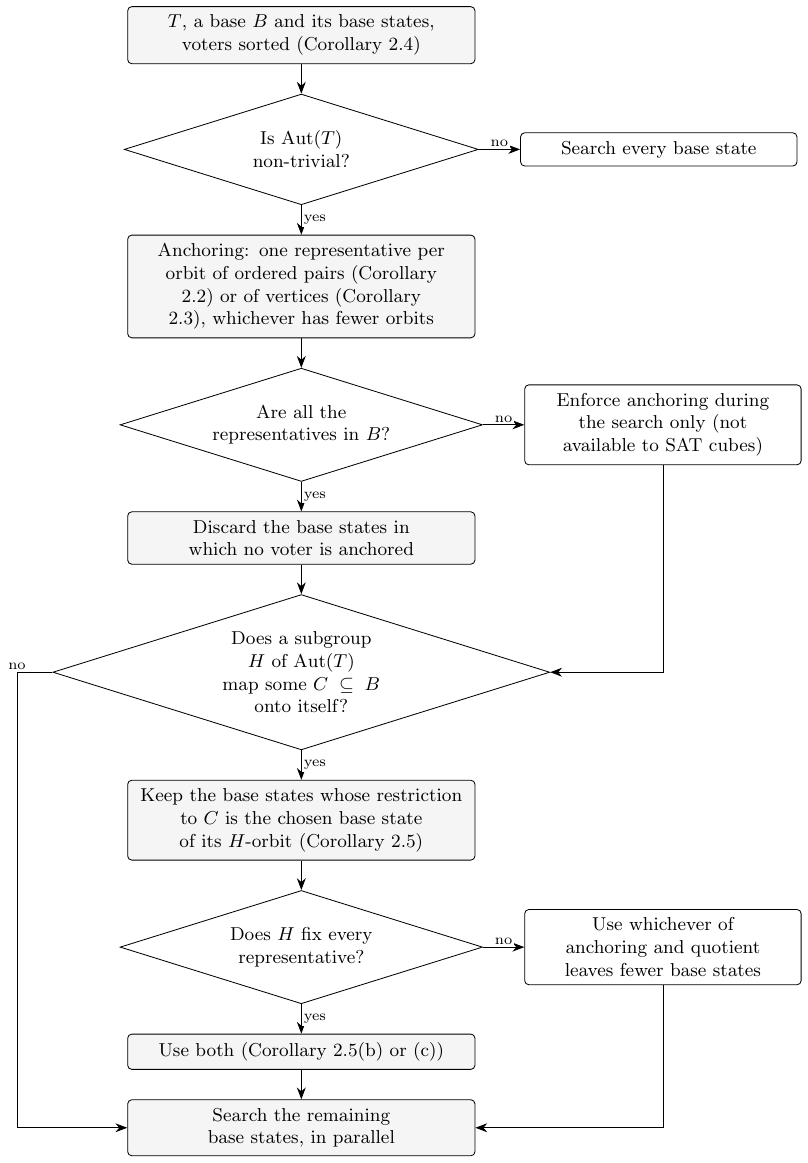}
\caption{Which symmetry reductions apply to a tournament $T$ and a base $B$. Anchoring by Corollary 2.2
or Corollary 2.3 can discard base states only when every representative lies in $B$, since a base
state reveals the orders on $B$ alone; otherwise the depth-first search enforces it while vertices are
inserted, and the SAT route cannot use it. The quotient of Corollary 2.5 combines with
anchoring only when $H$ fixes every representative.}
\end{figure}

\subsubsection{2.2 The search in five
steps}\label{the-search-in-five-steps}

The search extends base states. A base state fixes how the \(k\) voters
rank \(B\); the search inserts the remaining \(n - |B|\) vertices into
those rankings one at a time, backtracking as soon as the arcs already
decided contradict \(T\). A base state extended to \(k\) full rankings
yields a witness; one that cannot be is eliminated; and \(T\) fails to
be \(k\)-inducible exactly when every base state is eliminated. Given
\(T\) and the margin regime, the search has five steps:

\begin{enumerate}
\def\labelenumi{\arabic{enumi}.}
\tightlist
\item
  \textbf{Pick a size.} Choose the base size \(b\), in practice \(5\) or
  \(6\).
\item
  \textbf{Pick a subtournament.} Choose a set \(B\) of \(b\) vertices.
  Only \(T|_B\) matters for what follows, so the choice can and should
  be optimized quickly (see Appendix A.4), since the base-state count
  varies by more than two orders of magnitude across the \(12\)
  isomorphism classes at \(b = 5\).
\item
  \textbf{Find its base states.} Enumerate the \(k\)-tuples of linear
  orders of \(B\) whose supports agree with \(T|_B\), counted up to
  voter permutation by Corollary 2.4.
\item
  \textbf{Determine the usable symmetries.} Compute \(\mathrm{Aut}(T)\)
  with \texttt{nauty} and its orbits, then impose the ordered-pair
  anchoring of Corollary 2.2, the vertex anchoring of Corollary 2.3, or
  neither, discarding the base states no anchored witness can use. If a
  subgroup \(H\) maps \(B\), or a subset \(C\) of it, onto itself, also
  keep only the base states above one representative of each \(H\)-orbit
  (Corollary 2.5), checking that \(H\) fixes the anchoring
  representatives (Figure 2).
\item
  \textbf{Run to completion, in parallel across base states.} The
  verdict is positive as soon as any base state yields a witness. It is
  negative only when \emph{every} live base state has been searched to
  exhaustion.
\end{enumerate}

Figure A.1 in Appendix A.3 shows all five steps on \(P_7\) at \(k = 3\).
Steps 1--4 are fast and user-chosen; step 5 is a parallel computation.
The same steps describe the SAT route, with step 5 replaced by handing
each live base state to a CDCL solver as a \emph{cube} (the partial
assignment fixing the voters' rankings of \(B\)), so the two share their
decomposition and their symmetry reductions and differ only in what
refutes a leaf in the search tree.

\subsection{3. Results}\label{results}

The two regimes we explore are \emph{unrestricted} (or plain majority)
inducibility, and \emph{unit margin}. The former is the standard notion
of inducibility, and the latter asks that every arc be carried by
exactly three voters against two. Only an unrestricted refutation bounds
\(N(5)\). However, the lemma below shows that for the Paley family and
most tournaments derived from it, the unrestricted inducibility is
equivalent to inducibility with margin at most \(3\) (i.e.~support at
most 4).

We also use unit margin as a fast filter for unrestricted inducibility.
A unit-margin witness is in particular a majority witness, and the
unit-margin search is much faster (Appendix A.5), so when the question
is only whether a tournament is \(5\)-inducible we search at unit margin
first and at plain majority only if no unit-margin witness exists. The
order-\(12\) analysis of Appendix C works this way, and so do several of
the family searches on \(21\) vertices (Appendix G). A tournament that
fails the filter but has a majority witness is then a unit-margin
obstruction found as a by-product, and several of those reported below
arose this way.

\textbf{Lemma 3.1 (3-cycle bound).} \emph{Let \(u \to v \to w \to u\) be
a directed triangle of \(T\) and let \(\pi_1,\dots,\pi_k\) be any linear
orders with \(k\) odd. Then}
\[c(u\to v) + c(v \to w) + c(w \to u) \le 2k.\] \emph{Consequently, in
any majority witness, no arc lying in a directed triangle has unanimous
support \(k\).}

\emph{Proof.} A linear order cannot rank \(u\) above \(v\), \(v\) above
\(w\) and \(w\) above \(u\) simultaneously, so each voter agrees with at
most two of the three arcs; summing over the \(k\) voters gives the
inequality. If \(c(u \to v) = k\), the other two supports sum to at most
\(k\), contradicting the majority requirement that each be at least
\((k+1)/2\). \(\blacksquare\)

For \(k = 3\) the second conclusion is Proposition 1 of Milosz, Hamel
and Pierrot (2018), with a transitivity argument; we need it for every
odd \(k\), and in the summed form above.

\subsubsection{3.1 Paley(23) and Paley(31) are not
5-inducible}\label{paley23-and-paley31-are-not-5-inducible}

\textbf{Theorem 3.2.} \emph{\(P_{23}\) is not \(5\)-inducible.
Consequently \(N(5) \le 23\).}

The proof is a search that runs to completion. The base is
\(\{1,2,3,6,12\}\), which induces the regular tournament on \(5\)
vertices and gives \(8{,}031\) base states. They are reduced to
\(2{,}591\) by Corollary 2.2, by requiring one voter to rank either
\((1,3)\) or \((1,6)\) as the first pair. None of them produces a
witness. This statement is also formally certified in Appendix B.1.

Read as a statement about voting, it says that a five-member panel
cannot produce the collective preference \(P_{23}\) on twenty-three
alternatives: no assignment of five rankings to five voters, out of the
\(23!^{5}\) available, has \(P_{23}\) as its pairwise-majority relation.
It is the first such statement for an agenda of fewer than forty-three
alternatives, and the first for any tournament whose non-realizability a
general-purpose solver had been unable to settle.

The same method settles a larger case: \(P_{31}\) is not
\(5\)-inducible, over \(21{,}009\) base states with a different base,
reduced to \(4{,}007\) by Corollary 2.2. Appendix A.5 shows how the
runtime scales in \(q\), and Appendix A.6 tabulates the cost of every
computation in this paper. Settling \(P_{31}\) shows that our search
decides \(31\)-vertex instances on a single laptop within a day, which
the SAT and ILP solvers we tried did not.

\subsubsection{3.2 Eight tournaments on 22 vertices that five voters
cannot
produce}\label{eight-tournaments-on-22-vertices-that-five-voters-cannot-produce}

\(P_{23}\) is one of \(37\) doubly regular tournaments on \(23\)
vertices, which the catalogue of McKay (2026) lists up to isomorphism.
We write \(\Delta_i\) for the \(i\)-th tournament in that list, with its
vertices \(v_1, \dots, v_{23}\) numbered in the catalogue's own
labelling, and \(\Delta_i - v_j\) for the tournament left when vertex
\(v_j\) is deleted. Every one of the \(37\) is beyond five voters, at
unit margin and unrestricted alike (Section 3.7), so the doubly regular
tournaments on \(23\) vertices are a natural place to look for a smaller
obstruction one vertex down.

\textbf{Theorem 3.3.} \emph{The tournaments \(\Delta_1 - v_{1}\),
\(\Delta_1 - v_{2}\), \(\Delta_1 - v_{10}\), \(\Delta_6 - v_{5}\),
\(\Delta_{11} - v_{2}\), \(\Delta_{11} - v_{5}\),
\(\Delta_{11} - v_{20}\) and \(\Delta_{28} - v_{5}\), on \(22\)
vertices, are not \(5\)-inducible. In particular, no tournament obtained
from \(\Delta_1\) by deleting one vertex is \(5\)-inducible.
Consequently \(N(5) \le 22\).}

The second statement follows from the first because the automorphism
group of \(\Delta_1\), of order \(55\), has three vertex orbits, of
sizes \(11\), \(11\) and \(1\), which contain \(v_1\), \(v_2\) and
\(v_{10}\) respectively. The eight are pairwise non-isomorphic, and none
is isomorphic to the converse of another. Each has out-degrees \(10\)
and \(11\), eleven vertices of each, and every arc lies in at least five
directed triangles, so by the \(3\)-cycle bound no arc of a witness
could be unanimous.

All eight share an outline inherited from the doubly regular tournament
they come from. Deleting a vertex \(v\) of a doubly regular tournament
on \(23\) vertices leaves its \(11\) out-neighbours and its \(11\)
in-neighbours. Each of the two sets induces a regular tournament on
\(11\) vertices, and the \(121\) arcs between them are such that every
out-neighbour of \(v\) beats \(6\) in-neighbours and every in-neighbour
beats \(5\) out-neighbours. The obstructions differ in the two halves
and in how they are joined. In \(\Delta_1 - v_{10}\) and
\(\Delta_{11} - v_{20}\) both halves are \(P_{11}\). In
\(\Delta_1 - v_{1}\), \(\Delta_1 - v_{2}\), \(\Delta_{11} - v_{2}\) and
\(\Delta_{11} - v_{5}\) the two halves are again isomorphic to each
other and to each other's converse, but each has an automorphism group
of order \(5\) rather than \(55\). In \(\Delta_{28} - v_{5}\) the two
halves are converses of each other, each with no non-trivial
automorphism. In \(\Delta_6 - v_{5}\) the halves are neither isomorphic
nor converse, and again neither has a non-trivial automorphism. Appendix
H gives the adjacency matrices of all eight.

The halves suggest a definition. We call a vertex \(v\) of a tournament
\(T\) a \emph{bisector} if \(T[N^+(v)]\) is isomorphic to \(T[N^-(v)]\)
or to its converse, an \emph{isomorphic} or a \emph{converse} bisector
respectively; a vertex can be both. A tournament is \emph{bisectable} if
it has a bisector or arises from a tournament with one by deleting a
bisector, and \emph{non-bisectable} otherwise. A bisector has equal in-
and out-degree, so the first case needs odd order and the second even
order, and a tournament of even order is bisectable exactly when its
vertices split into two halves of equal size that induce isomorphic or
converse tournaments. Bisectable tournaments are much easier to
enumerate than the others, since one tournament on half the vertices
fixes both halves. Every doubly regular tournament on at most \(23\)
vertices is bisectable, whereas \(118\) of the \(722\) on \(27\)
vertices are not. On \(23\) vertices, \(P_{23}\) is self-converse and
the other \(36\) form \(18\) pairs of converses. Every vertex is a
bisector only in \(P_{23}\) and in \(\Delta_1\) and its converse
\(\Delta_{35}\). The halves at a vertex of \(P_{23}\) are the same
tournament at every vertex, and it is a half of no vertex of any other;
in \(\Delta_1\) the halves at \(v_{10}\) are \(P_{11}\), while those at
\(v_1\) and \(v_2\) are not. Our obstructions cover every case. In
\(\Delta_1 - v_{10}\) and the five deletions with an automorphism of
order \(5\) the deleted vertex is a bisector of both types, and in
\(\Delta_{28} - v_{5}\) a converse bisector, so these seven are
bisectable; \(\Delta_1 - v_{10}\) and \(\Delta_{28} - v_{5}\) are
bisectable through a single split of their vertices, the one into their
two halves. The deleted vertex of \(\Delta_6\) is not a bisector, and no
other split works either, so \(\Delta_6 - v_{5}\) is non-bisectable.

The first is the most structured. Figure 1 shows \(\Delta_1 - v_{10}\)
as two copies of \(P_{11}\), on \(X = \{x_0, \dots, x_{10}\}\) and
\(Y = \{y_0, \dots, y_{10}\}\), joined by a ladder whose rungs all point
from \(Y\) to \(X\), and its automorphism group has order \(55\). Its
two vertex orbits are \(X\) and \(Y\). The automorphisms multiplying
every index by a non-zero square fix \(x_0\) and \(y_0\) and form a
subgroup \(H\) of order \(5\), which maps the five out-neighbours of
\(x_0\) in \(X\), \(B = \{x_1, x_3, x_4, x_5, x_9\}\), onto themselves.
\(B\) induces the regular tournament on five vertices, so it has
\(8{,}031\) base states, which \(H\) groups into \(1{,}611\) orbits.
Corollary 2.5 therefore lets the search keep one base state per orbit,
and since \(H\) fixes \(x_0\) and \(y_0\), it can at the same time
anchor on them as the representatives of Corollary 2.3. The complete
search over the \(1{,}611\) chosen base states finds no witness.

The five further deletions keep part of this symmetry. Deleting a vertex
of \(\Delta_1\) from one of its orbits of size \(11\) leaves an
automorphism group of order \(5\); \(\Delta_{11}\) itself has an
automorphism group of order \(5\), which fixes \(v_2\), \(v_5\) and
\(v_{20}\), so its three deletions \(\Delta_{11} - v_{2}\),
\(\Delta_{11} - v_{5}\) and \(\Delta_{11} - v_{20}\) keep it. In each of
the five, this group maps a five-vertex base inducing the regular
tournament on five vertices onto itself, so Corollary 2.5 again reduces
the \(8{,}031\) base states to \(1{,}611\) chosen base states, and the
complete search over them finds no witness. These five refutations are
not formally certified; Appendix A.6 gives their cost.

The refutation of \(\Delta_1 - v_{10}\) is also formally certified
(Appendix B.1). The certificate uses the quotient of Corollary 2.5 but
no anchoring, with a different base,
\(B' = \{y_1, y_3, y_4, y_5, y_9\}\), the five in-neighbours of \(y_0\)
in \(Y\). \(H\) maps \(B'\) onto itself as well, and \(B'\) also induces
the regular tournament on five vertices, so the certificate again keeps
\(1{,}611\) of the \(8{,}031\) base states.

\(\Delta_6 - v_{5}\) and \(\Delta_{28} - v_{5}\) have no non-trivial
automorphism, so none of Corollary 2.2, Corollary 2.3 and Corollary 2.5
applies and every base state has to be searched. Each proof is a
complete search over all \(8{,}031\) base states of a five-vertex base
inducing the regular tournament on five vertices, none of which extends
to a witness. The refutation of \(\Delta_6 - v_{5}\) is also formally
certified, by the same pipeline as those of \(P_{19}\) and \(P_{23}\)
(Appendix B.1).

Read as a statement about committees, the theorem moves the threshold at
which five voters stop being enough by one more alternative: on an
agenda of twenty-two alternatives there are already patterns of pairwise
majority that no five-member panel can produce. All eight patterns sit
one alternative away from perfect balance.

All eight obstructions are \emph{vertex-critical}: deleting any one of
the \(22\) vertices of any of them leaves a \(5\)-inducible tournament,
so none yields a \(21\)-vertex obstruction by a further deletion. For
\(\Delta_1 - v_{10}\) this takes only two deletions, one from each
vertex orbit. Appendix G summarizes the wider search below \(22\)
vertices, which has found no obstruction there. Section 3.3 adds a ninth
obstruction on \(22\) vertices, obtained from \(\Delta_1 - v_{10}\) by
reversing one arc.

\subsubsection{3.3 Criticality across the Paley
family}\label{criticality-across-the-paley-family}

Arc-criticality implies vertex-criticality: reversing an arc incident to
\(v\) leaves \(T - v\) unchanged, and the restriction of a witness to
\(V \setminus \{v\}\) is a witness for \(T - v\). Each cell below
therefore records only the \emph{strongest} property established, at
each of the two margins. ``Not vertex-critical'' therefore also rules
out arc-criticality.

\begin{longtable}[]{@{}lll@{}}
\caption{The strongest criticality property established for each Paley
tournament, at each margin.}\tabularnewline*
\toprule\noalign{}
\(q\) & unit margin & unrestricted \\*
\midrule\noalign{}
\endfirsthead
\toprule\noalign{}
\(q\) & unit margin & unrestricted \\*
\midrule\noalign{}
\endhead
\bottomrule\noalign{}
\endlastfoot
19 & \textbf{arc-critical} & \(5\)-inducible, so criticality does not
apply \\*
23 & not vertex-critical & \textbf{arc-critical} (Section 3.6) \\*
31 & not vertex-critical & \textbf{not vertex-critical} (Section 3.5) \\*
43 & not vertex-critical & \textbf{not vertex-critical} (Section 3.5) \\*
\end{longtable}

\(P_{19}\) is the only member critical at unit margin, and it is
critical in the strongest sense: reversing any one arc makes it
unit-margin inducible, which by the implication also gives
vertex-criticality. At the other end, \(P_{31}\) and \(P_{43}\) fail to
be vertex-critical at \emph{both} margins. The interesting cell is
\(q = 23\): it is \emph{vertex-critical unrestricted}, but \emph{not}
vertex-critical at unit margin. In other words, \(P_{23} -\,v\) is
5-inducible, but not with unit margin, analogous to the following
result.

The eight \(22\)-vertex obstructions of Section 3.2 are vertex-critical
unrestricted. Under its automorphism group, the \(231\) arcs of
\(\Delta_1 - v_{10}\) fall into five orbits: the arcs inside \(X\), the
arcs inside \(Y\), the rungs \(y_i \to x_i\), the other arcs from \(Y\)
to \(X\), and the arcs from \(X\) to \(Y\). Reversing a rung yields a
\(5\)-inducible tournament, but reversing one of the other arcs from
\(Y\) to \(X\), \(y_j \to x_i\) with \(j - i \in \mathrm{QR}\), does
not. The tournament it yields has no non-trivial automorphism, and a
complete search over all \(8{,}031\) base states of a five-vertex base
inducing the regular tournament on five vertices finds no witness. So
\(\Delta_1 - v_{10}\) is not arc-critical, and reversing one of its arcs
gives a ninth obstruction on \(22\) vertices, not isomorphic to any of
the eight of Section 3.2 or to the converse of any. This ninth
obstruction is vertex-critical too. Deleting either endpoint of the
reversed arc leaves \(\Delta_1 - v_{10}\) minus a vertex, which is
\(5\)-inducible, and each of the other \(20\) deletions has an explicit
witness. Whether reversing an arc of the remaining three orbits restores
\(5\)-inducibility is open, and so is the arc-criticality of the other
seven obstructions of Section 3.2.

\subsubsection{3.4 Paley(19) is 5-inducible, but not with unit
margin}\label{paley19-is-5-inducible-but-not-with-unit-margin}

That \(P_{19}\) is \(5\)-inducible is due to Bachmeier et al.~(2019),
who computed \(\dim(Q_{19}) = 5\). Our search recovers it independently
and contributes the negative half: it is \emph{not} \(5\)-inducible with
unit margin. This is a complete refutation over all \(2{,}200\) base
states of the base \(\{1,2,3,4,6\}\), and it is the second statement we
certify formally in Appendix B.1. These two facts supply the separation
in Section 1: no arc of \(P_{19}\) can be unanimous, by the \(3\)-cycle
bound above, so it is \(5\)-inducible with margin at most \(3\) and not
with unit margin, so the inclusion of the unit-margin regime in margin
at most \(3\) is strict.

\(P_{19}\) is within reach of five voters, but not by bare three-to-two
majorities alone: any five rankings that realize it must settle at least
one pair by four votes to one.

Our four certified refutations, \(P_{23}\), \(\Delta_1 - v_{10}\) and
\(\Delta_6 - v_{5}\) at unrestricted margin in Sections 3.1 and 3.2 and
\(P_{19}\) at unit margin, are established formally by cube-and-conquer
computations in which every leaf is refuted by CaDiCaL 2.0.0 (Biere et
al.~2024; Pollitt et al.~2023) in LRAT and rechecked by
\texttt{lrat-trim} (Pollitt et al.~2023), a separate program that shares
no code with the solver, with the solver's own proof checking
deliberately disabled. All four are complete on \emph{both} halves (the
leaves and the claim that the leaves are exhaustive). Those for
\(P_{19}\) and \(P_{23}\) reduce to the same two human-checked results,
Corollaries 2.2 and 2.4; that for \(\Delta_1 - v_{10}\) to Corollaries
2.4 and 2.5; and that for \(\Delta_6 - v_{5}\), which has no non-trivial
automorphism, to Corollary 2.4 alone (see Appendix B).

\subsubsection{3.5 Paley(31) and Paley(43) are not
vertex-critical}\label{paley31-and-paley43-are-not-vertex-critical}

Vertex-criticality is the natural minimality notion for an
\emph{obstruction}, a tournament that is not \(5\)-inducible, and
\(P_{43}\), the first explicit non-\(5\)-inducible tournament in our
companion paper, is the natural candidate.

\textbf{Theorem 3.4.} \emph{Neither \(P_{31}\) nor \(P_{43}\) is
vertex-critical for 5-inducibility.}

\textbf{The orbit break.} For any \(q \equiv 3 \pmod 4\),
\(\mathrm{Aut}(P_{q} - v)\) is \emph{exactly} the stabilizer of \(v\) in
\(\mathrm{Aut}(P_{q})\). One inclusion is immediate; for the other,
deleting \(v\) removes an out-arc from every vertex of \(N^{-}(v)\) and
from no vertex of \(N^{+}(v)\), so \(P_q - v\) has just two out-degrees,
\((q-1)/2\) on \(N^{+}(v)\) and \((q-1)/2 - 1\) on \(N^{-}(v)\), and an
automorphism preserves out-degree, hence those two sets, hence extends
to \(P_q\) by fixing \(v\). Translating so that \(v = 0\), that
stabilizer is \(\{x \mapsto ax : a \in \mathrm{QR}\}\), cyclic of order
\((q-1)/2\), and it splits the remaining \(q - 1\) vertices into exactly
two orbits of size \((q-1)/2\), namely \(\mathrm{QR}\) and its
complement. Two representatives, one from each, therefore meet every
orbit, and the vertex anchoring of Corollary 2.3 applies with those two.

Each sweep runs over all \(8{,}031\) base states of a five-vertex base
inducing the regular tournament on five vertices: \(\{1,2,3,4,13\}\) in
\(P_{31} - v\) and \(\{1,2,3,4,11\}\) in \(P_{43} - v\). The two counts
coincide because, as in Appendix A.2, the number of base states depends
only on the induced subtournament and on the margin, and not on \(n\).

Every arc of \(P_{q} - v\) lies in at least \((q-3)/4\) directed
triangles, so no arc of a majority witness could be unanimous, and
margin \(\le 3\) is not a restriction at all. The refutation at margin
\(\le 3\) is therefore an unrestricted refutation. We use the same
approach for both \(q = 31\) and \(q = 43\).\\
We thus independently reconfirm the main result of our companion paper,
that \(P_{43}\) is not \(5\)-inducible: if it were, so would be
\(P_{43} - v\), by deleting \(v\) from every voter's order.

\subsubsection{3.6 Vertex and arc criticality for
Paley(23)}\label{vertex-and-arc-criticality-for-paley23}

\textbf{Theorem 3.5.} \emph{\(P_{23}\) is arc-critical: reversing any
single arc yields a \(5\)-inducible tournament. Consequently it is
vertex-critical.}

Let \(T^{e}\) denote \(P_{23}\) with the arc \((1,2)\) reversed; since
\(\mathrm{Aut}(P_{23})\) is transitive on arcs, any two single-arc
reversals of \(P_{23}\) are isomorphic, so we may pick \((1,2)\) without
loss of generality. The computation is a scan of the \(8{,}031\) base
states of \(T^{e}\), but now \(T^{e}\) is rigid, so Corollary 2.2 and
Corollary 2.3 do not apply. The witness we found was verified
independently of the search that produced it, by two checks on the five
orders: each is a permutation of the \(23\) vertices, and the majority
tournament they induce agrees with \(T^{e}\) in every one of the \(253\)
arcs and therefore differs from \(P_{23}\) in exactly \(e\).

\textbf{Vertex- and arc-criticality come apart, as far as they can.}
Arc-criticality implies vertex-criticality, so it is natural to ask
whether the two coincide. They do not. Among the \(110\)
vertex-transitive tournaments on \(21\) vertices, four are unit-margin
obstructions, and all four are vertex-critical. However, arc-criticality
takes all three of its possible values among those same four: two of
them have every arc critical, one has exactly half, and one has none at
all, so no statement of the form ``this tournament is or is not
arc-critical'' captures the situation (full details in Appendix F).

\subsubsection{3.7 Other structured tournament
families}\label{other-structured-tournament-families}

Beyond the Paley family we swept the structured families catalogued by
McKay (2026) exhaustively for \(n < 23\). Every verdict is negative for
the bound, so we only report them briefly.

First, \emph{the vertex-transitive family contains no majority
obstruction at all for \(n < 23\)}: every one is \(5\)-inducible, all at
unit margin except \(P_{19}\) and four of the \(110\) on \(21\)
vertices, which have majority witnesses. Since only a majority
refutation can move \(N(5)\), that family is closed as a source of
candidates below \(23\), and the closure survives one-arc perturbation,
since reversing any single arc of the four unit-margin obstructions, one
representative per arc orbit, again leaves a majority-inducible
tournament in all \(40\) cases.

Second, \emph{both} doubly regular tournaments on \(19\) vertices,
\(P_{19}\) and the one other of that order, are unit-margin obstructions
and both are majority-inducible, so the separation that makes \(P_{19}\)
interesting is not a property of \(P_{19}\) in particular. Nor is its
criticality: the second one is \emph{arc-critical} at unit margin too.
Its automorphism group has order \(3\), so its \(171\) arcs fall into
\(57\) orbits of size \(3\), and reversing the representative of any one
of them yields a tournament with a unit-margin witness: \(57\) orbits
and \(57\) witnesses. Arc-criticality at unit margin is therefore shared
by both doubly regular tournaments of order \(19\).

Third, every regular tournament on at most \(15\) vertices is inducible
at unit margin, over all \(R_{15} \approx 1.8 \times 10^{10}\) of those
on \(15\) vertices, and with the instance counts summing to OEIS
A096368(7) (OEIS Foundation Inc.~2026c). Likewise, every self-converse
tournament on \(13\) vertices is inducible at unit margin, over all
\(S_{13} \approx 9.5 \times 10^{7}\) of them. With the order-\(12\)
analysis of Appendix C, which covers every smaller order by inheritance,
that settles every self-converse tournament on at most \(13\) vertices.

Fourth, the order \(23\) is complete for both families. Every
vertex-transitive tournament on \(23\) vertices other than \(P_{23}\) is
\(5\)-inducible, while none of the \(37\) doubly regular ones is, at
unit margin or unrestricted. Up to converse the \(37\) form \(19\)
classes, and each was settled by a complete search (Appendix A.6).

Taken together with the Paley results above and the families of Appendix
G, these searches turn up \(58\) tournaments, up to isomorphism and
converse, that five voters can produce but not with unit margin: \(2\)
doubly regular ones at order \(19\); at order \(21\), \(3\)
vertex-transitive ones (the four of Section 3.6, two of which are
converses of each other) and \(42\) near-doubly regular ones; and \(11\)
circulant ones at order \(23\). Several of them were found as a
by-product of the unit-margin filter described at the start of this
section.

\textbf{One family needs no search.} A tournament is \emph{locally
transitive} if every in- and out-neighbourhood induces a transitive
subtournament. Every such tournament is \(k\)-inducible at unit margin
for every odd \(k \ge 3\), by an explicit construction rather than a
computation; Appendix E provides it. This is the one structural family
for which the inducibility question is settled outright.

\subsection{4. Discussion and open
problems}\label{discussion-and-open-problems}

\subsubsection{4.1 Other applications of incremental depth-first search
with
MRV}\label{other-applications-of-incremental-depth-first-search-with-mrv}

Subject to the caveats of Appendix A.6, the shape of the engine may
transfer: a placement search in which each decision \emph{refines},
rather than recomputes, the candidate sets of all remaining objects,
with a dynamic minimum-remaining-values (MRV) order and an externally
discharged symmetry break. The conditions are specific: the extensions
of a partial solution must factor as a product that shrinks
monotonically under each decision, and the symmetry group must be
computable; a general-purpose solver may win if they do not apply.

Neither device is new in isolation, and the closest antecedent is
Knuth's Algorithm X with dancing links (Knuth 2000). Its usual
column-choice heuristic (the column with the fewest remaining ones,
which Knuth credits to Golomb and Baumert) is our MRV order, and its
link surgery does what our refinement does: the state is altered in
place and put back on backtrack, so a node costs what changed rather
than what the whole state contains. The constraint is what differs.
Exact cover partitions, whereas an arc of a tournament imposes a
threshold, at least three of five voters, so the search here is not an
exact cover instance and dancing links does not apply to it as it
stands.

Two design lessons do seem to generalize. First, the incremental
refinement, not the heuristic order, carried most of the speedup here, a
factor of \(110\) against a factor of a few. Second, a symmetry break
should be \emph{parameterized} by its representatives rather than
hard-coded, so that its correctness obligation becomes a separate
computation instead of an assumption buried in the search; Section 3.5
is a case where discharging that obligation required an explicit orbit
calculation that the engine could not have performed for itself.

\subsubsection{4.2 Conjectures and open
problems}\label{conjectures-and-open-problems}

\textbf{Question 1.} \emph{Is \(N(5) = 22\), that is, is every
tournament on \(21\) vertices \(5\)-inducible?}

The first version of this paper (arXiv:2609.13924v1), whose upper bound
was \(N(5) \le 23\), conjectured that \(N(5) = 23\). The obstructions on
\(22\) vertices of Section 3.2 refute that conjecture, so we pose the
question one order lower instead.

Both ends of \(13 \le N(5) \le 22\) are constructive and improve on
those of our companion paper. The unit-margin obstructions at \(n = 19\)
and \(21\) cannot move the upper one, since every one of them is
majority-inducible. The evidence below \(22\) comes mostly from the
descendants of the doubly regular tournaments on \(23\) vertices
(Appendix G): every tournament on \(20\) vertices obtained from one of
them by deleting three vertices is \(5\)-inducible, and so is every one
on \(21\) vertices below any of the nine \(22\)-vertex obstructions.

Every class we searched below \(22\) vertices was chosen because
symmetry, a construction, or descent from a known obstruction made it
searchable, and Appendix G lists them all. These classes cluster around
the doubly regular tournaments, where every obstruction we know of lies.
What they show is that no obstruction on \(21\) vertices lies
\emph{among them}; they say little about the roughly
\(3 \times 10^{43}\) tournaments on \(21\) vertices as a whole.

\textbf{Conjecture 2.} \emph{\(P_{q}\) is not \(5\)-inducible for any
eligible prime \(q \ge 23\), i.e.~\(q \equiv 3 \pmod 4\).}

It holds at \(q = 23\), \(31\) and \(43\), and only finitely many \(q\)
are in doubt: \(\alpha^{*}(P_q) = \mathrm{MAS}(P_q)/C\) tends to
\(1/2\), because Paley tournaments are quasi-random in the sense of
Chung and Graham (1991, Section 6), and one of their equivalent
quasi-random properties (\(P_9\)) is that every linear order of the
vertices agrees with \((1/2 + o(1))\binom{q}{2}\) arcs. Hence the
predictability eventually falls below \(3/5\), past which the necessary
condition of Section 1.2 refutes \(5\)-inducibility with no search at
all. What is missing is a bound on where that range ends; the values in
Section 1.2 fall too irregularly to extrapolate one.

\textbf{Open problem 1 (vertex- and arc-criticality).} Whether \(P_{q}\)
minus \emph{two} vertices is \(5\)-inducible, equivalently whether
\(P_{q}\) minus one vertex is vertex-critical, is open at \(q = 31\) and
\(q = 43\).

\textbf{Open problem 2 (the margin hierarchy).} The first of unit margin
\(\subseteq\) margin \(\le 3\) \(\subseteq\) unrestricted is strict
(Sections 3.4 and 3.7); whether the second ever is, that is, whether
some tournament is inducible by five voters but not at margin \(\le 3\),
is open. Lemma 3.1 forces such a tournament to have an arc in \emph{no}
directed triangle, which rules out every regular tournament. Conjecture
8.1 of our companion paper states that the inclusion is strict at every
odd \(k > 3\).

\textbf{Conjecture 3.} \emph{For every odd \(k\), every \(k\)-inducible
tournament is \((k+2)\)-inducible with unit margin.}

Conjecture 3 extends Conjecture 8.1 of our companion paper, which
concerns the classes for a single \(k\), to classes across different
\(k\). Adding an order together with its reverse raises every support by
one, so \(k\)-inducibility with margin at most \(M\) implies
\((k+2)\)-inducibility with margin at most \(M\). Conjecture 3
strengthens this to unit margin, and with the inclusions within each
\(k\) it would nest every class at a lower \(k\) inside every class at a
higher \(k\), whatever the margins. It trivially holds at \(k = 1\),
since a transitive tournament is the majority of an order taken twice
and its reverse once, with every arc carried two against one. At
\(k = 3\), Appendix C shows that every tournament on at most \(12\)
vertices is \(5\)-inducible with unit margin, so a counterexample needs
at least \(13\) vertices. At \(k = 5\), both doubly regular tournaments
on \(19\) vertices, which are \(5\)-inducible but not with unit margin,
are \(7\)-inducible with unit margin.

\subsection{5. Reproducibility}\label{reproducibility}

\subsubsection{5.1 The repository}\label{the-repository}

The code, data and verdict files are at
\url{https://github.com/Leonardini/TournamentsBeyond5Voters}. Its
\texttt{CLAIMS.md} carries one row per statement made here, naming the
artifact that holds the verdict and the command that re-derives it, and
it ends with an explicit list of what the package does not establish. A
permanent archive of the commit corresponding to the accepted version
will be deposited on acceptance and its DOI recorded here.

The repository contains the search engine as a single C translation
unit, the SAT encoding and cube tooling in Python, the certification
driver, the independent witness verifiers, the tournaments as bit
strings, and a verdict record for every result reported here. Each
verdict records the exact command line, the base and break used, the
coverage audit, the runtime, and the two root hashes for the formally
certified results.

\subsubsection{5.2 What must and what may not
replicate}\label{what-must-and-what-may-not-replicate}

The certified computations publish two root hashes, and only one of them
is portable.

\textbf{ROOT (CNF)} commits to the SHA-256 of every cube's CNF, in cube
order. These files are generated by the repository's own code from the
tournament and the base, and are reproducible on any machine: a third
party who regenerates the cube set \emph{must} obtain this exact value.
It is the meaningful cross-check, because it proves they solved the same
problems.

\textbf{ROOT (proofs)} additionally commits to the LRAT proof bytes.
These are \emph{not} portable, as they depend on the exact SAT solver
and version used. A mismatch here is not evidence of an error.

We have run that check: on the cluster, under a different compiler and
CaDiCaL build, both refutations gave matching roots at \(70.9\) and
\(567.6\) cluster core-hours against the \(25.1\) and \(236.0\) of
Appendix A.6, and both coverage instances rebuilt to their committed CNF
hashes and were refuted again.

\textbf{Search runtimes: nodes replicate, hours do not.} A three-way
version of that distinction applies to the uncertified sweeps.
\emph{Verdicts} replicate under any correct implementation, which is the
claim the paper makes. \emph{Node counts} replicate exactly, but only
with five choices held fixed:

\begin{enumerate}
\def\labelenumi{\arabic{enumi}.}
\tightlist
\item
  the same tie-breaking rule in the variable order and in the domain
  refinement; ties are common and a different rule explores a different
  tree of the same size class but not the same size;
\item
  the same vertex numbering of the tournament;
\item
  the same base size and base;
\item
  the same resulting set of base states, which follows from (2) and (3);
\item
  the same margin regime.
\end{enumerate}

The engine prints all of these except (2) in a header on every run
(margin, variable order, tie-break mode, symmetry break, base size,
base, base arc pattern and base-state count), so a run log is a
self-describing fingerprint. Condition (2) is the tournament file
itself, which is why the repository stores tournaments rather than
regenerating them from their constructions. \emph{Core-hours} replicate
only in order of magnitude; Appendix A.5 measures a factor of \(2\) to
\(5\) per core between the two machines we used.

\subsection{Declaration of competing
interest}\label{declaration-of-competing-interest}

The authors declare that they have no known competing financial
interests or personal relationships that could have appeared to
influence the work reported in this paper.

\subsection{Funding}\label{funding}

LC acknowledges funding from the MRC Centre for Global Infectious
Disease Analysis (reference MR/X020258/1), funded by the UK Medical
Research Council (MRC). This UK funded award is carried out in the frame
of the Global Health EDCTP3 Joint Undertaking.

\subsection{CRediT authorship contribution
statement}\label{credit-authorship-contribution-statement}

\textbf{Leonid Chindelevitch:} Conceptualization, Methodology, Software,
Formal analysis, Investigation, Writing -- original draft, Writing --
review and editing. \textbf{Ararat Harutyunyan:} Conceptualization,
Formal analysis, Writing -- review and editing.

\subsection{Data availability}\label{data-availability}

The code, data and verdict files supporting the findings of this study
are at \url{https://github.com/Leonardini/TournamentsBeyond5Voters},
described in Section 5.

\subsection{Declaration of generative AI and AI-assisted technologies in
the manuscript preparation
process}\label{declaration-of-generative-ai-and-ai-assisted-technologies-in-the-manuscript-preparation-process}

During the preparation of this work the authors used Claude (Anthropic)
for exploratory reasoning, implementation assistance, and drafting and
editing. After using this tool, the authors reviewed and edited the
content as needed and take full responsibility for the content of the
published article.

\subsection{Acknowledgements}\label{acknowledgements}

This work was granted access to the HPC resources of IDRIS under the
allocation 2025-AD010616623R1 made by the Grand Équipement National de
Calcul Intensif (GENCI).

\subsection{References}\label{references}

Alspach B, Tabib C (1982) A note on the number of 4-circuits in a
tournament. \emph{Annals of Discrete Mathematics} 12:13--19

Bachmeier G, Brandt F, Geist C, Harrenstein P, Kardel K, Peters D,
Seedig HG (2019) \(k\)-Majority digraphs and the hardness of voting with
a constant number of voters. \emph{Journal of Computer and System
Sciences} 105:130--157

Biere A, Faller T, Fazekas K, Fleury M, Froleyks N, Pollitt F (2024)
CaDiCaL 2.0. In: \emph{Computer Aided Verification (CAV 2024)}, LNCS
14681, Springer, pp 133--152. \url{doi:10.1007/978-3-031-65627-9_7}

Brakensiek J, Heule M, Mackey J, Narváez D (2020) The resolution of
Keller's conjecture. In: \emph{Automated Reasoning (IJCAR 2020)}, LNCS
12166, Springer, pp 48--65

Brouwer AE (1980) The enumeration of locally transitive tournaments.
Report ZW 138/80, Mathematisch Centrum, Afdeling Zuivere Wiskunde,
Amsterdam

Chindelevitch L, Harutyunyan A (2026) Tournaments determined by three
and five voters. Submitted to \emph{Mathematical Social Sciences}
jointly with this paper

Chung FRK, Graham RL (1991) Quasi-random tournaments. \emph{Journal of
Graph Theory} 15(2):173--198

Cruz-Filipe L, Heule MJH, Hunt WA Jr, Kaufmann M, Schneider-Kamp P
(2017) Efficient certified RAT verification. In: \emph{Automated
Deduction (CADE 2017)}, LNCS 10395, Springer, pp 220--236

Cruz-Filipe L, Marques-Silva J, Schneider-Kamp P (2019) Formally
verifying the solution to the Boolean Pythagorean triples problem.
\emph{Journal of Automated Reasoning} 63(3):695--722.
\url{doi:10.1007/s10817-018-9490-4}

Dummit DS, Foote RM (2004) \emph{Abstract Algebra}, 3rd edn. Wiley,
Hoboken, NJ

Dushnik B, Miller EW (1941) Partially ordered sets. \emph{American
Journal of Mathematics} 63(3):600--610

Eggermont C, Hurkens C, Woeginger GJ (2013) Realizing small tournaments
through few permutations. \emph{Acta Cybernetica} 21(2):267--271

Heule MJH (2018) Schur number five. In: \emph{Proceedings of the AAAI
Conference on Artificial Intelligence (AAAI 2018)}, pp 6598--6606

Heule MJH, Kullmann O, Marek VW (2016) Solving and verifying the Boolean
Pythagorean triples problem via cube-and-conquer. In: \emph{Theory and
Applications of Satisfiability Testing (SAT 2016)}, LNCS 9710, Springer,
pp 228--245

Heule MJH, Kullmann O, Wieringa S, Biere A (2012) Cube and conquer:
guiding CDCL SAT solvers by lookaheads. In: \emph{Haifa Verification
Conference (HVC 2011)}, LNCS 7261, Springer, pp 50--65

Knuth DE (2000) Dancing links. In: Davies J, Roscoe B, Woodcock J (eds)
\emph{Millennial Perspectives in Computer Science}, Palgrave, pp
187--214. arXiv:cs/0011047

McGarvey DC (1953) A theorem on the construction of voting paradoxes.
\emph{Econometrica} 21(4):608--610

McKay BD (2026) Combinatorial data: digraphs (tournament catalogues).
Accessed 10 August 2026.
\url{https://users.cecs.anu.edu.au/~bdm/data/digraphs.html}

McKay BD, Piperno A (2014) Practical graph isomorphism, II.
\emph{Journal of Symbolic Computation} 60:94--112

Milosz R, Hamel S, Pierrot A (2018) Median of 3 permutations, 3-cycles
and 3-hitting set problem. In: \emph{Proceedings of IWOCA 2018}, LNCS
10979, Springer, pp 224--236. \url{doi:10.1007/978-3-319-94667-2_19}

OEIS Foundation Inc.~(2026a) Entry A000568, Number of outcomes of
unlabeled \(n\)-team round-robin tournaments. The On-Line Encyclopedia
of Integer Sequences. \url{https://oeis.org/A000568}

OEIS Foundation Inc.~(2026b) Entry A002785, Number of self-complementary
oriented graphs with \(n\) nodes. The On-Line Encyclopedia of Integer
Sequences. \url{https://oeis.org/A002785}

OEIS Foundation Inc.~(2026c) Entry A096368, Number of unlabeled regular
tournaments with \(2n+1\) nodes. The On-Line Encyclopedia of Integer
Sequences. \url{https://oeis.org/A096368}

Pollitt F, Fleury M, Biere A (2023) Faster LRAT checking than solving
with CaDiCaL. In: \emph{Theory and Applications of Satisfiability
Testing (SAT 2023)}, LIPIcs 271, pp 21:1--21:12.
\url{doi:10.4230/LIPIcs.SAT.2023.21}

Shepardson D, Tovey CA (2009) Smallest tournaments not realizable by
\(2/3\)-majority voting. \emph{Social Choice and Welfare}
33(3):495--503. \url{doi:10.1007/s00355-009-0375-7}

\subsection{Appendix A. The search
engine}\label{appendix-a.-the-search-engine}

\setcounter{table}{0}\renewcommand{\thetable}{A.\arabic{table}}

\setcounter{figure}{0}\renewcommand{\thefigure}{A.\arabic{figure}}

Integer programming and satisfiability are both central to this project.
Every optimum reported in the companion paper is certified by an integer
or linear program, and the integer program of A.1 below is that paper's
formulation B.1 with antisymmetry substituted out; the only
machine-checked theorem in this paper is a SAT computation from end to
end, with CaDiCaL (Biere et al.~2024; Pollitt et al.~2023) emitting LRAT
proofs that \texttt{lrat-trim} (Pollitt et al.~2023) checks. The claim
here is narrow: on this one decision problem, at this one range of
sizes, neither tool reaches the answer while a search built around the
problem's own structure does, and the same SAT machinery is then
indispensable for turning that answer into a proof.

\subsubsection{A.1 Integer programming and SAT are both too
slow}\label{a.1-integer-programming-and-sat-are-both-too-slow}

The natural integer program has one binary \(x_{v,ij}\) per voter \(v\)
and pair \(i < j\), meaning that \(v\) ranks \(i\) above \(j\), with
\(x_{v,ji}\) read as \(1 - x_{v,ij}\) so that antisymmetry is
structural. Each voter is forced to be a linear order by forbidding both
orientations of every \(3\)-cycle,
\[0 \;\le\; x_{v,ij} + x_{v,jl} - x_{v,il} \;\le\; 1 \qquad (i < j < l),\]
and the profile is forced to induce \(T\) by one majority row per arc,
\(\sum_{v} x_{v,ij} \ge (k+1)/2\). This is a pure feasibility problem.
At \(q = 23\), \(k = 5\) it has \(1{,}265\) binaries, \(17{,}710\)
transitivity rows and \(253\) majority rows.

The obstruction is not a deficiency of the solver. Let us set
\(x_{v,ij} = 3/5\) on every arc of \(T\) and \(2/5\) on every non-arc,
the same for all voters. Every cyclic triple then reads \(4/5\) and
every transitive one \(3/5\), both interior to \([0,1]\), while every
majority row is tight at exactly \(3\). The relaxation is therefore
feasible, and since the problem carries no objective, LP feasibility
yields no bound whatever: branch-and-bound is left with nothing to prune
by except infeasibility discovered deep in the tree. That fractional
point is also invariant under both symmetry groups acting on the problem
(the \(k!\) voter relabellings and \(\mathrm{Aut}(T)\)), so it supplies
no branching direction either, and orbital branching recovers only the
\(k!\) factor. CPLEX and Gurobi both failed to close instances at
\(n = 19\) within a day.

The satisfiability encoding is \emph{dissent-Boolean}: one variable per
voter \(v\) and arc \(i \to j\) of \(T\), true when \(v\) ranks \(j\)
above \(i\) and so dissents from that arc. Each arc bounds its dissents,
at most \((k-1)/2\) of them for majority and exactly \((k-1)/2\) for
unit margin, and each voter's dissent set is forced to leave a linear
order triple by triple: a tournament is transitive exactly when it has
no directed \(3\)-cycle, so it is enough to forbid, for every voter and
every triple, the two dissent patterns that would make that triple
cyclic. For \(k = 5\) that is \(5\binom{n}{2}\) variables, \(15\)
clauses per arc at unit margin and \(10\) per triple. It is compact and
propagates well locally, and it is what the certified refutations of
Appendix B are built on.

Table 4 of Bachmeier et al.~(2019) reports their
\texttt{Sat-Check-k-Majority} on uniform random tournaments, tabulating
only entries whose average stays under 20 seconds. The \(k = 5\) column
reads

\begin{longtable}[]{@{}llllllll@{}}
\caption{Average time of Sat-Check-k-Majority at \(k = 5\) on uniform
random tournaments (Bachmeier et al.~2019, Table 4).}\tabularnewline*
\toprule\noalign{}
\(n\) & 18 & 19 & 20 & 21 & 22 & 23 & 24 \\*
\midrule\noalign{}
\endfirsthead
\toprule\noalign{}
\(n\) & 18 & 19 & 20 & 21 & 22 & 23 & 24 \\*
\midrule\noalign{}
\endhead
\bottomrule\noalign{}
\endlastfoot
seconds & \(0.23\) & \(0.35\) & \(0.54\) & \(5.87\) & \(11.07\) &
\(18.95\) & --- \\*
\end{longtable}

so on \emph{random} instances the method leaves the table between
\(n = 23\) and \(n = 24\), roughly doubling per added vertex. Random
tournaments are the easy case: no automorphisms to rediscover and, being
far from extremal, many witnesses. On the structured instance of the
same order, \(P_{23}\), the same solver did not terminate within a
cumulative six weeks. The gap between \(19\) seconds and six weeks at
equal \(n\) is the cost of symmetry.

A better encoding and an explicit symmetry break both help substantially
and are still not enough (results not shown). What does work is
supplying the solver with the decomposition itself: cube-and-conquer
over base states (Appendix A.2), which is how the refutations of
Appendix B are certified. A CDCL solver must otherwise discover that
decomposition for itself, and cannot exploit \(\mathrm{Aut}(T)\) at all
without an explicit break in the encoding.

\subsubsection{A.2 Base states}\label{a.2-base-states}

Every search starts by fixing how the voters rank a handful of vertices
and then looks for ways to add the remaining ones. We choose a
\emph{base} \(B \subseteq V\) of \(b\) vertices (we use \(b = 5\) or
\(b = 6\)), and write \(T|_B\) for the sub-tournament induced on \(B\).
When \(B\) is a base, a \emph{base state of \(B\)} (we omit \(B\) when
the context makes it clear) is a \(k\)-voter profile over \(B\) that is
consistent with \(T|_B\), i.e.~ \(k\) orderings of \(B\) whose
majorities reproduce \(T|_B\) at the margin being asked for. Base states
differing only by a relabelling of the voters are counted once (see
Corollary 2.4).

Each base state of \(B\) defines a self-contained subproblem: the
rankings of \(B\) are given, and the question is whether the remaining
\(n - |B|\) vertices can be inserted so that every arc of \(T\) has the
right support, so the search splits into independent components across
base states. How many there are depends only on \(|B|\), on \(T|_B\) and
on the margin regime, but \emph{not} on \(n\); thus, with the
five-vertex bases we use on the Paley tournaments, which induce regular
sub-tournaments, there are \(8{,}031\) of them in both \(P_{23}\) and
\(P_{47}\).

\subsubsection{A.3 Depth-first placement
search}\label{a.3-depth-first-placement-search}

The search maintains a partial profile: a set \(S \subseteq V\) of
\emph{placed} vertices together with, for each voter, a linear order of
\(S\). It begins from a base state, which places \(B\), and extends by
inserting one unplaced vertex at a time. Inserting \(v\) means choosing,
for each voter independently, a slot among the \(|S| + 1\) available;
the \emph{domain} \(D(v \mid S)\) is the set of \(k\)-tuples of slots
consistent with every arc between \(v\) and \(S\). A partial profile is
extendable only if every unplaced vertex has a non-empty domain, and
domains shrink monotonically as \(S\) grows, reducing the search to a
constraint-propagation problem. Two decisions in the pseudocode below
enable its efficiency (see Appendix A.4 for more details). First, the
search always inserts an unplaced vertex of currently smallest domain
(the minimum-remaining-values, or MRV, rule), the domain sizes being
recomputed at every node rather than fixed once per base state; second,
a child's domains are refined from its parent's. It also shows where the
symmetry reductions of Section 2.1 enter (Appendix A.4).

\begin{figure}[!htbp]
\begin{verbatim}
ALGORITHM 1.  Placement search for k-inducibility of T

Input :  tournament T on V;  k voters;  margin (majority or unit);  base B
         Q: chosen base states of B    // sorted (Cor 2.4), one per H-orbit (Cor 2.5)
         R: anchoring representatives  // vertices (Cor 2.3), pairs (Cor 2.2), or none
Output:  a witness profile, or a refutation

MAIN
  for each base state in Q  (a lex-canonical assignment of the base B to k orders):
      S <- B
      D(v | S) <- fresh enumeration, for every v not in S
      if any D(v | S) is empty:  continue     // this base state is dead
      if DFS(S) succeeds:        return its witness
  return "not k-inducible"                    // no base state extends

DFS(S)
  if S = V:  return the profile               // all vertices placed
  if every vertex of R is placed and no voter is anchored:
      return failure                          // Cor 2.2/2.3; at the root if R is in B
  v* <- argmin  |D(v | S)|   over v not in S  // MRV, recomputed at EVERY node
        ties broken toward  max | |out(v) & S| - |in(v) & S| |
  for each slot tuple p in D(v* | S):
      insert v* at p in each of the k orders;  S' <- S u {v*}
      for each w not in S':
          D(w | S') <- REFINE( D(w | S), p, w, v* )
          if D(w | S') is empty:  undo;  continue with the next p
      if DFS(S') succeeds:  return its witness
      undo                                    // arena pointer reset
  return failure

REFINE(D, q, w, u)                            // q = the slot tuple just used for u
  D' <- {}
  for each p in D:
      lift p to p' voter by voter:
          p_i <  q_i  ->  p'_i = p_i          // w stays before u
          p_i >  q_i  ->  p'_i = p_i + 1      // w stays after u, shifted by the insertion
          p_i == q_i  ->  BOTH are legal, so the tuple splits in two
      keep those lifts whose support  c_w(u) = #{ i : w before u }
      meets the requirement of the single new arc between w and u
  return D'
\end{verbatim}
\end{figure}

\begin{figure}[!t]
\centering
\includegraphics[width=\linewidth]{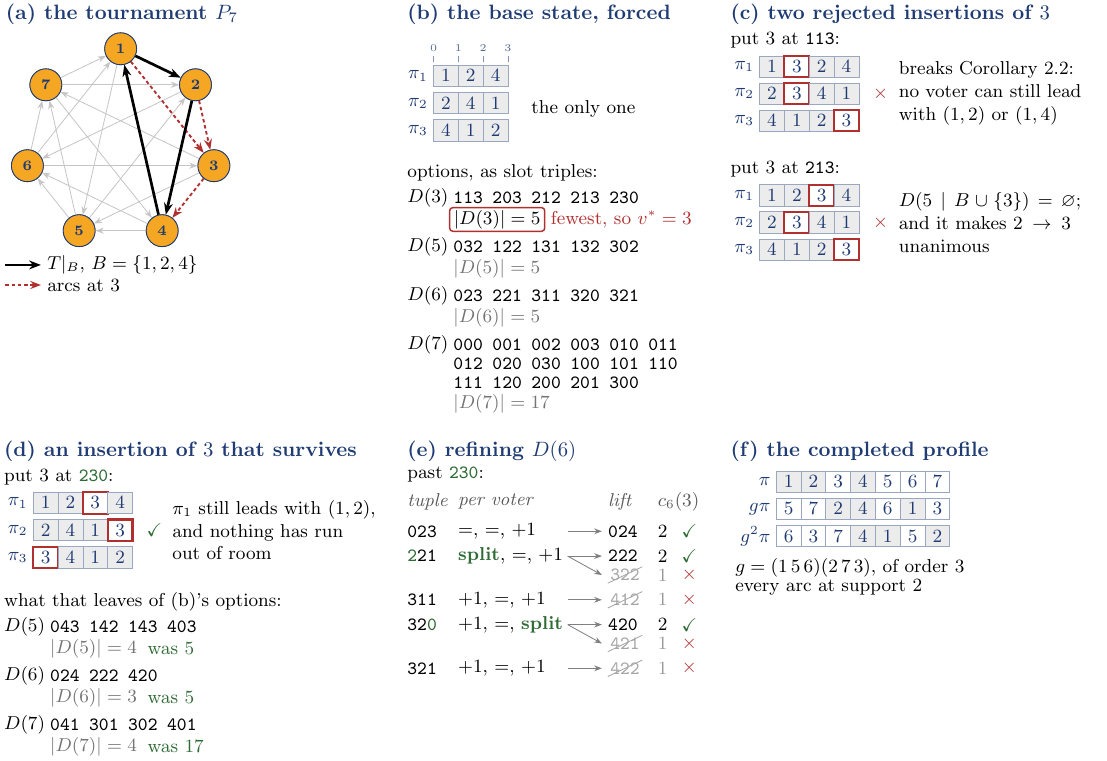}
\caption{Algorithm 1 on $P_7$ at $k = 3$, which is $3$-inducible, so the run ends in a witness. 
A \emph{slot triple} is one gap per voter, of the $64$ available, and a vertex's domain is
those triples its arcs to $B$ admit. Here the base is $B = \mathrm{QR}$, the squares modulo 7, 
and Corollary 2.2 guarantees that at least one voter must rank 1 first and either 2 or 4 second. 
Ranking 4 second forces the other two voters to $241$, the only order getting both $4 \to 1$ and
$2 \to 4$ right, and $1 \to 2$ is then left with support $1$; so the base state in (b) is the only one.
\textbf{(a)} The base $T|_B$ on $B = \{1,2,4\}$ in bold; dashed, the arcs at the first vertex
inserted. \textbf{(b)} Steps 2--4: the base state is forced, and MRV takes the smallest domain.
\textbf{(c)} Two insertions the search rejects, one by Corollary 2.2 and one by leaving $5$ no slot at all.
\textbf{(d)} One that survives, and what it leaves of (b). \textbf{(e)} REFINE lifts each parent
tuple past the slot just used (unchanged below it, shifted above it, split where the two
coincide, which is what green marks), and the one new arc then keeps or kills each lift.
\textbf{(f)} The witness, every arc at support exactly $2$. Without Corollary 2.2 the search returns
all $7$ witnesses, a single orbit of $\mathrm{Aut}(P_7)$, of order $21$, so the profile is unique
up to symmetry, as Shepardson and Tovey (2009) report; $\pi$ generates the other two under the $g$
shown, the map $x \mapsto 2x + 4$ of $\mathbb{F}_7$.}
\end{figure}

\subsubsection{A.4 What makes the search
fast}\label{a.4-what-makes-the-search-fast}

Four choices make the search of Appendix A.3 fast, and the second is the
one that made \(n = 19\) and \(n = 23\) reasonable to compute.

\textbf{A dynamic variable order.} At every node we insert the unplaced
vertex minimizing \(|D(v \mid S)|\), breaking ties toward the vertex
whose arcs to \(S\) are most imbalanced. The important part is that the
order is recomputed at every node rather than fixed per base state: on
Paley instances a static order is not competitive, because how
constrained a vertex is depends strongly on which part of the tournament
has already been placed.

\textbf{Incremental domain refinement.} Recomputing \(D(v \mid S)\) from
scratch at each node is the dominant cost of a naive implementation;
instead the child's domain is derived from the parent's by the three
cases in \texttt{REFINE} above, after which the single new arc filters
the lifted tuples by their support. Child domains are therefore
refinements of parent domains, computed in time proportional to the
surviving tuples rather than to the whole slot space, and domains live
in a stack-disciplined arena indexed by (depth, vertex) so that
backtracking is a pointer reset. Like for like this is worth a factor of
about \(110\) in wall time.

\textbf{Symmetry breaking.} We reduce the base states using the voter
relabellings (Corollary 2.4) and the orbits of a subgroup of
\(\mathrm{Aut}(T)\) mapping the base onto itself (Corollary 2.5), and we
restrict the search using the anchors (Corollary 2.2 or 2.3), tested at
every node once all representatives are placed; Section 2.1 and Figure 2
give the details.

\textbf{Early termination and caching by arc pattern.} A quick per-voter
agreement bound, obtained by counting the agreement between voter \(i\)
and the arcs between \(v\) and \(S\) two ways, proves \(D(v \mid S)\)
empty in \(O(k|S|)\) time without enumerating it. And the per-base-state
work is keyed by the \emph{arc pattern} of \(T|_B\) rather than by \(B\)
itself, so base-state counts, which depend only on that pattern, are
reused across every copy of the same base class and across every \(q\).

\textbf{Choosing the base.} The base set is not a free parameter, and
the fewest live base states is not the fastest choice: at \(q = 31\) the
class with \(4{,}007\) live states beat the class with \(2{,}537\),
whose states are individually slower, so we rank the twelve \(5\)-vertex
classes by the product of the live count and a sampled time per state.

\subsubsection{\texorpdfstring{A.5 How the runtime scales in
\(q\)}{A.5 How the runtime scales in q}}\label{a.5-how-the-runtime-scales-in-q}

\textbf{A note on \(q = 27\).} The series below includes \(q = 27\), the
one prime power among the orders we computed with. Since \(27 = 3^3\),
\(P_{27}\) must be built over \(\mathbb{F}_{27}\) and not over
\(\mathbb{Z}/27\). \(P_{27}\) is the unique most symmetric doubly
regular tournament on \(27\) vertices, with
\(|\mathrm{Aut}(P_{27})| = 1053 = 3q(q-1)/2\), because the field
automorphisms contribute (in general \(|\mathrm{Aut}(P_q)| = q(q-1)e/2\)
when \(q = p^e\)). It is still transitive on arcs and on non-arcs, which
is what Corollary 2.2 needs. We ran \(q = 27\) to completion and it is
not \(5\)-inducible; since that does not bear on \(N(5)\) we report it
here only as a runtime datapoint.

On a fixed base decomposition the total runtime \emph{decreases} as the
tournament grows, and so does the node count: a larger Paley tournament
is \emph{more} constrained, so the search is refuted nearer the root and
the tree above each base state is smaller rather than larger. The work
per node moves the other way, rising monotonically from \(10.8\) to
\(57.4\) microseconds across the four runs below, because a larger
tournament carries larger domains and each refinement step costs more.
The node count falls faster than the per-node cost rises, which is what
makes the total fall. The full runs:

\begin{longtable}[]{@{}
  >{\raggedright\arraybackslash}p{(\linewidth - 8\tabcolsep) * \real{0.0864}}
  >{\raggedright\arraybackslash}p{(\linewidth - 8\tabcolsep) * \real{0.2469}}
  >{\raggedright\arraybackslash}p{(\linewidth - 8\tabcolsep) * \real{0.1728}}
  >{\raggedright\arraybackslash}p{(\linewidth - 8\tabcolsep) * \real{0.1481}}
  >{\raggedright\arraybackslash}p{(\linewidth - 8\tabcolsep) * \real{0.3457}}@{}}
\caption{The four full Paley sweeps.}\tabularnewline*
\toprule\noalign{}
\begin{minipage}[b]{\linewidth}\raggedright
\(q\)
\end{minipage} & \begin{minipage}[b]{\linewidth}\raggedright
live base states
\end{minipage} & \begin{minipage}[b]{\linewidth}\raggedright
core-hours
\end{minipage} & \begin{minipage}[b]{\linewidth}\raggedright
nodes
\end{minipage} & \begin{minipage}[b]{\linewidth}\raggedright
seconds per live base state
\end{minipage} \\*
\midrule\noalign{}
\endfirsthead
\toprule\noalign{}
\begin{minipage}[b]{\linewidth}\raggedright
\(q\)
\end{minipage} & \begin{minipage}[b]{\linewidth}\raggedright
live base states
\end{minipage} & \begin{minipage}[b]{\linewidth}\raggedright
core-hours
\end{minipage} & \begin{minipage}[b]{\linewidth}\raggedright
nodes
\end{minipage} & \begin{minipage}[b]{\linewidth}\raggedright
seconds per live base state
\end{minipage} \\*
\midrule\noalign{}
\endhead
\bottomrule\noalign{}
\endlastfoot
23 & \(2{,}591\) & \(34.03\) & \(1.14 \times 10^{10}\) & \(47.3\) \\*
27 & \(2{,}537\) & \(31.04\) & \(6.22 \times 10^{9}\) & \(44.0\) \\*
31 & \(4{,}007\) & \(26.89\) & \(3.43 \times 10^{9}\) & \(24.2\) \\*
43 & \(2{,}537\) & \(22.87\) & \(1.43 \times 10^{9}\) & \(32.5\) \\*
\end{longtable}

\(P_{43}\) is decided in \emph{less time than \(P_{27}\)}, and explores
\(4.3\) times fewer nodes to do it. The last column is the third divided
by the second, so only the core-hours and the live base states are
measured. Just two of these four runs share a decomposition, \(q = 27\)
and \(q = 43\), on the same five-vertex base class and the same
\(2{,}537\) live base states; \(q = 23\) and \(q = 31\) each used their
own best configuration, and \(q = 31\)'s is the cheaper class of
Appendix A.4, which is why its per-state figure is the lowest of the
four. The core-hours fall monotonically in \(q\) either way. By
contrast, \(P_{43}\) minus a vertex takes \(185.5\) core-hours, far more
than \(P_{43}\) itself, as vertex deletion shrinks the automorphism
group.

Additionally, the search gets faster as the margin is tightened. On the
identical tournament (\(P_{43} -\,v\)), engine and \(5\)-vertex base
class, there is a factor of about \(90\). Requiring every arc to be at
unit margin admits \(2{,}200\) base states where majority admits
\(8{,}031\); additionally, under exact margin the last voter is
\emph{forced}, since the support accumulated over the first \(k-1\)
voters determines what the last must contribute, so a whole level of the
search collapses.

\textbf{Two machines, and why their runtimes are not comparable.} Some
computations ran on a cluster rather than the laptop. Measured across
six independent experiments, the cluster ran at \(2\) to \(5\) times the
laptop's runtime per unit of work.

We are not claiming that our search beats ILP or SAT solvers in general.
Our engine is narrow: it exploits the specific facts that a partial
profile's extensions form a product of per-voter slot choices, that this
product refines monotonically, and that \(\mathrm{Aut}(T)\) is large.
For the specific Paley instances in this paper that the general-purpose
methods did not settle for us, the search settles in hours on a single
laptop.

\subsubsection{A.6 Runtimes for every
computation}\label{a.6-runtimes-for-every-computation}

All runs in this manuscript were performed on a single laptop (Apple
M-series, 14 cores, 24 GB RAM), using at most 11 cores concurrently,
with the following exceptions, which used a compute cluster: the census
of regular tournaments on \(15\) vertices, the self-converse census of
order \(13\), the order-\(12\) census of Appendix C, six of the
\(22\)-vertex sweeps of Section 3.2, part of the certificate for
\(\Delta_1 - v_{10}\), \(16\) of the \(19\) unrestricted sweeps of
doubly regular tournaments on \(23\) vertices in Section 3.7, and most
of the search of Appendix G. Runtimes are reported in \emph{core-hours},
the workers' running times summed; cluster core-hours are not comparable
with laptop ones (Appendix A.5), so we convert every cluster runtime to
\emph{laptop-equivalent} core-hours, as follows. Every core-hour figure
in this paper is on the laptop's scale, except the two cluster figures
that Section 5 marks as such. The certificate for \(\Delta_1 - v_{10}\)
ran partly on each machine. Both machines solved \(133\) of its groups
of five cubes, and on those the cluster took on average about \(2.1\)
times as long as the laptop, between \(1.6\) and \(3.1\) times for
individual groups; we divide its cluster time by that average. The
factor is an average slowdown rather than an exact exchange rate, so the
converted total is approximate. The searches of \(\Delta_1 - v_{1}\),
\(\Delta_1 - v_{2}\), \(\Delta_{11} - v_{5}\) and
\(\Delta_{11} - v_{20}\) also ran on the cluster, wholly or in part, and
we convert them with a factor measured on these searches: \(355\) base
states of \(\Delta_{11} - v_{5}\) were searched on both machines, and
the cluster took \(3.55\) times as long in total. The factor differs
from the certificate's because the two computations run different
programs. We apply the same factor to the other searches of this kind
that ran on the cluster, the sweeps of \(\Delta_6 - v_{5}\) and
\(\Delta_{28} - v_{5}\) and of \(16\) of the doubly regular tournaments
on \(23\) vertices, for which no paired measurement exists. The
self-converse census of order \(13\) ran partly on each machine, and
there the cluster took \(5.13\) times as long per tournament, so we
divide its cluster time by that factor. The censuses of orders \(12\)
and \(15\) ran on the cluster alone, so we measured their factors
directly: we reran a random sample of their parts on the laptop, \(8\)
of order \(12\) and \(12\) of order \(15\), with the same scripts, and
the cluster took \(3.85\) and \(4.28\) times as long respectively.

\begin{table}[!htbp]
\centering
\caption{The runtime of every computation in this paper.}
\begin{tabular}{@{}
  >{\raggedright\arraybackslash}p{(\linewidth - 6\tabcolsep) * \real{0.5729}}
  >{\raggedright\arraybackslash}p{(\linewidth - 6\tabcolsep) * \real{0.1458}}
  >{\raggedright\arraybackslash}p{(\linewidth - 6\tabcolsep) * \real{0.1562}}
  >{\raggedleft\arraybackslash}p{(\linewidth - 6\tabcolsep) * \real{0.1250}}@{}}
\toprule\noalign{}
\begin{minipage}[b]{\linewidth}\raggedright
computation
\end{minipage} & \begin{minipage}[b]{\linewidth}\raggedright
margin
\end{minipage} & \begin{minipage}[b]{\linewidth}\raggedright
verdict
\end{minipage} & \begin{minipage}[b]{\linewidth}\raggedleft
core-hours
\end{minipage} \\
\midrule\noalign{}

\(P_{19}\), certified refutation & unit & not inducible & \(25.1\) \\
\(P_{23}\), certified refutation & unrestricted & not inducible &
\(236.0\) \\
\(\Delta_1 - v_{10}\), sweep (Corollary 2.5) & unrestricted & not
inducible & \(96.9\) \\
\(\Delta_1 - v_{10}\), certified refutation & unrestricted & not
inducible & \(732\) \\
\(\Delta_6 - v_{5}\), certified refutation & unrestricted & not
inducible & \(1{,}292.7\) \\
\(\Delta_6 - v_{5}\), sweep & unrestricted & not inducible &
\(437.0\) \\
\(\Delta_{28} - v_{5}\), sweep & unrestricted & not inducible &
\(433.2\) \\
\(16\) doubly regular on \(23\) vertices & unrestricted & not inducible
& \(7{,}458.2\) \\
\(\Delta_1 - v_{1}\), sweep (Corollary 2.5) & unrestricted & not
inducible & \(136.8\) \\
\(\Delta_1 - v_{2}\), sweep (Corollary 2.5) & unrestricted & not
inducible & \(122.9\) \\
\(\Delta_{11} - v_{2}\), sweep (Corollary 2.5) & unrestricted & not
inducible & \(108.5\) \\
\(\Delta_{11} - v_{5}\), sweep (Corollary 2.5) & unrestricted & not
inducible & \(124.2\) \\
\(\Delta_{11} - v_{20}\), sweep (Corollary 2.5) & unrestricted & not
inducible & \(93.6\) \\
\(P_{31} -\,v\), sweep & unrestricted & not inducible & \(239.5\) \\
\(P_{23}\), one arc reversed & unrestricted & inducible & \(210.4\) \\
second doubly regular \(19\), all \(57\) arc reversals & unit & all
inducible & \(3.59\) \\
regular on \(15\) vertices & unit & all inducible & \(726.1\) \\
self-converse on \(13\) vertices & unit & all inducible & \(294.2\) \\
all on \(12\) vertices & unit & all inducible & \(2{,}615.6\) \\
\bottomrule\noalign{}
\end{tabular}
\end{table}

The four full Paley sweeps, \(q = 23\), \(27\), \(31\) and \(43\), are
tabulated with their node counts in Appendix A.5.

\subsection{Appendix B. The formal
certificates}\label{appendix-b.-the-formal-certificates}

\setcounter{table}{0}\renewcommand{\thetable}{B.\arabic{table}}

\subsubsection{B.1 The certified
refutations}\label{b.1-the-certified-refutations}

All four machine-checked results follow an established design.
\emph{Cube-and-conquer} is due to Heule, Kullmann, Wieringa and Biere
(2012): split a hard instance into many partial assignments
(\emph{cubes}), hand each to a CDCL solver, and let the splitting supply
the global structure the solver cannot find for itself. The standard of
\emph{proof} is the one Heule, Kullmann and Marek set for the Boolean
Pythagorean triples problem (Heule et al.~2016), later re-checked with a
formally verified checker (Cruz-Filipe et al.~2019), and Schur number
five (Heule 2018) and the Keller conjecture (Brakensiek et al.~2020)
after it: every subproblem emits a machine-checkable refutation,
validated by an independent checker rather than trusted. We claim no
methodological novelty.

\textbf{What an LRAT proof is.} A solver's verdict of unsatisfiability
is not self-evidently checkable, so the solver writes a \emph{proof
log}: a file recording, in order, the clauses it derives, each of which
must follow from the input formula together with the clauses already
recorded, the last being the empty clause. A separate checker replays
that file and confirms every step, so what one trusts is the checker
rather than the solver. The available formats record essentially this
same derivation and differ mainly in how much work the replay costs. We
use LRAT (Cruz-Filipe et al.~2017), the format CaDiCaL emits natively
and \texttt{lrat-trim} consumes, in which each derived clause lists the
earlier clauses it follows from, so checking it needs no search (Pollitt
et al.~2023).

Two differences are consequences of our setting. Our cubes are
\emph{base states}, a combinatorial decomposition, which lets the
coverage of the cube set be certified separately (B.2) instead of
resting on a splitting heuristic. And where Heule et al.~(2016)
published a cube certificate from which its refutations can be
regenerated, we \emph{verify and discard}: each LRAT proof is checked,
hashed and deleted, and we only publish the hash value. The choice suits
the scale. Verifying the Boolean Pythagorean triples proof costs about
\(13{,}000\) core-hours to regenerate the refutations from the published
cube certificate and about \(16{,}000\) to check them (Heule et
al.~2016). Our certificates are orders of magnitude smaller: \(25.1\)
core-hours for \(P_{19}\), \(236.0\) for \(P_{23}\), \(1{,}292.7\) for
\(\Delta_6 - v_{5}\) and about \(732\) for \(\Delta_1 - v_{10}\) cover
solving \emph{and} checking. Anyone replicating the computation can
compare the portable root values (Section 5.2).

For \(P_{19}\) at unit margin the instance is the dissent-Boolean
encoding of Appendix A.1 at \(q = 19\), \(k = 5\): \(855\) variables and
\(12{,}255\) clauses. The cubes are the \(142{,}251\) base states of the
\(6\)-vertex base \(\{1,2,3,4,5,12\}\), of which \(22{,}876\) survive
the arc-orbit break, and each is handed whole to CaDiCaL with
\texttt{-\/-lrat=true\ -\/-checkproof=0}. Each LRAT proof is checked by
\texttt{lrat-trim}. The runtime was \(2.64\) hours on ten cores. Its
coverage half is the same construction as the one below, at
\(143{,}032\) clauses, and solves in \(7.56\) seconds.

For the \(P_{23}\) refutation of Section 3.1, the base is
\(\{1,2,3,4,6,7\}\) with \(3{,}414{,}729\) base states, of which
\(343{,}896\) survive the arc-orbit and non-arc-orbit breaks; every one
is unsatisfiable, with an LRAT proof independently verified by
\texttt{lrat-trim}. The runtime was \(21.72\) hours on eleven cores. The
coverage half of the proof is a single unsatisfiable instance of \(350\)
constraint clauses, \(356\) symmetry-breaking clauses and
\(3{,}414{,}729\) negated cubes, one per base state, since coverage must
see \emph{every} base state whereas the orbit breaks apply to the search
half only. It solves in just under \(20\) minutes on one core and emits
a \(2.05\) GiB LRAT proof.

For the \(\Delta_6 - v_{5}\) refutation of Section 3.2 the instance is
the dissent-Boolean encoding at \(n = 22\) and plain majority:
\(1{,}155\) variables and \(17{,}710\) clauses. The tournament has no
non-trivial automorphism, so there is no orbit break, and all
\(742{,}374\) base states of the six-vertex base \(\{1,2,3,4,13,19\}\)
(in the labelling of \(\Delta_6\)) are cubes; every one is
unsatisfiable, with an LRAT proof independently verified by
\texttt{lrat-trim}. The runtime was about \(118\) hours on eleven cores.
The coverage half is the same construction as for \(P_{23}\), with
\(742{,}374\) negated cubes, and solves in \(82.8\) seconds on one core.
This refutation is the most expensive of the four, although its
tournament is smaller than \(P_{23}\): with no automorphisms there is
nothing to break, so every base state reaches the solver, against one in
ten for \(P_{23}\).

For the \(\Delta_1 - v_{10}\) refutation of Section 3.2 the instance is
the same encoding as for \(\Delta_6 - v_{5}\), at \(n = 22\) and plain
majority. The cubes are the \(1{,}611\) orbit representatives under
\(H\) of the \(8{,}031\) base states of the five-vertex base
\(B' = \{13, 15, 16, 20, 22\}\) (in the labelling of \(\Delta_1\)), one
per orbit, as Corollary 2.5 licenses; every one is unsatisfiable, with
an LRAT proof independently verified by \texttt{lrat-trim}. The cubes
were solved in groups of five, part on the laptop and part on the
cluster, at about \(732\) core-hours (Appendix A.6). The coverage half
is the same construction as for \(P_{23}\), with \(8{,}031\) negated
cubes among its \(8{,}467\) clauses over \(90\) variables, and solves in
\(0.08\) seconds on one core. A cube here fixes the voters' orders on
five vertices rather than six, so the cubes are about \(460\) times
fewer than for \(\Delta_6 - v_{5}\), and each is correspondingly harder.

\subsubsection{B.2 What the certificates do and do not
establish}\label{b.2-what-the-certificates-do-and-do-not-establish}

The trust chain has four links:

\begin{longtable}[]{@{}
  >{\raggedright\arraybackslash}p{(\linewidth - 2\tabcolsep) * \real{0.5273}}
  >{\raggedright\arraybackslash}p{(\linewidth - 2\tabcolsep) * \real{0.4727}}@{}}
\caption{The four links of the trust chain.}\tabularnewline*
\toprule\noalign{}
\begin{minipage}[b]{\linewidth}\raggedright
link
\end{minipage} & \begin{minipage}[b]{\linewidth}\raggedright
status
\end{minipage} \\*
\midrule\noalign{}
\endfirsthead
\toprule\noalign{}
\begin{minipage}[b]{\linewidth}\raggedright
link
\end{minipage} & \begin{minipage}[b]{\linewidth}\raggedright
status
\end{minipage} \\*
\midrule\noalign{}
\endhead
\bottomrule\noalign{}
\endlastfoot
every live cube is unsatisfiable & machine-checked: CaDiCaL, then
\texttt{lrat-trim} \\*
the cubes cover every lex-canonical base assignment & machine-checked:
LRAT, verified by \texttt{lrat-trim} \\*
(L1) orbit anchoring loses no generality & human-checked: Corollary
2.2 \\*
(L2) lex-ordering the voters loses no generality & human-checked:
Corollary 2.4 \\*
(L3) one base state per orbit loses no generality & human-checked:
Corollary 2.5 \\*
\end{longtable}

The human-checked links are proved in Section 2. Each certificate uses
only some of them: \(P_{19}\) and \(P_{23}\) use (L1) and (L2),
\(\Delta_1 - v_{10}\) uses (L2) and (L3), and \(\Delta_6 - v_{5}\) uses
(L2) alone. Corollary 2.2 is what licenses solving only the \emph{live}
cubes, in its arc/non-arc form: \(\mathrm{Aut}(P_q)\) is transitive on
arcs and on non-arcs, so a voter's top pair may be assumed to be a fixed
arc or a fixed non-arc. Corollary 2.4 licenses the lex chain in the
coverage computation. Corollary 2.5 licenses solving one base state per
orbit of \(H\) for \(\Delta_1 - v_{10}\), while the coverage computation
still runs over all \(8{,}031\) base states. \(\Delta_6 - v_{5}\) has no
automorphism to exploit, so neither orbit argument is used. These
results are human-checked links because nothing in the pipeline checks
them; the depth-first engine inherits them, but shares no implementation
with this one, so their agreement is evidence about the implementations,
not the results.

\subsection{Appendix C. The order-12 case
analysis}\label{appendix-c.-the-order-12-case-analysis}

\setcounter{table}{0}\renewcommand{\thetable}{C.\arabic{table}}

Every tournament on \(12\) vertices is \(5\)-inducible, so
\(N(5) \ge 13\). This appendix describes that computation. It is one of
the results in this paper that was not run on the laptop of Section 3:
it ran on a shared cluster, and its runtime is converted as Appendix A.6
describes.

\subsubsection{C.1 Reduction to one-vertex
extensions}\label{c.1-reduction-to-one-vertex-extensions}

A direct census is out of reach: \(D_{12} \approx 1.5 \times 10^{11}\)
classes, \(170\) times the order-\(11\) census of Chindelevitch and
Harutyunyan (2026). We use that census as the outer loop instead. Let us
write \(L + S\) for the tournament obtained from an order-\(11\)
tournament \(L\) by adding a vertex \(v\) whose out-neighbourhood is
\(S \subseteq V(L)\). Two observations simplify the analysis.

\textbf{(i)} Every order-\(12\) tournament is \(L + S\) for some
order-\(11\) \(L\) and some \(S\).

\textbf{(ii)} Restricting a profile that realizes \(L + S\) to \(V(L)\)
gives a profile that realizes \(L\).

By (i), enumerating all \(D_{11}\) order-\(11\) classes and settling all
\(2{,}048\) extensions of each settles every order-\(12\) tournament;
all of the former are \(5\)-inducible by the census of the companion
paper, which is what makes the outer loop available. By (ii), a profile
realizing \(L + S\) may be sought by \emph{inserting} \(v\) into a
profile realizing \(L\) with no loss of generality, since every profile
for \(L+S\) arises that way. That is what phase 1 exploits.

\subsubsection{C.2 Three phases}\label{c.2-three-phases}

The census runs in three phases, fastest first.

\textbf{Phase 1, the covering search.} Rather than \(2{,}048\) separate
instances per order-\(11\) tournament, one search enumerates profiles
realizing \(L\) \emph{at unit margin} and, for each, records every
out-neighbourhood obtainable by inserting \(v\) into its five orders:
\(v\)'s position in each order is a threshold, and the support of
\(v \to u\) is the number of orders that place \(v\) above \(u\), so one
traversal certifies many extensions at once. This is the engine of
Appendix A.3 with \(v\)'s placement read off rather than fixed in
advance. The margin makes this phase pay twice: it keeps the enumeration
of profiles for \(L\) small, and a unit-margin witness is in particular
a majority witness, so every extension the phase covers is settled in
both regimes at once. The phase is one-sided: an extension it covers is
certified \(5\)-inducible, but one left over is not refuted. This phase
settles every one of the \(2{,}048\) extensions of \(99.5\%\) of the
order-\(11\) tournaments.

\textbf{Phase 2, deciding what is left.} Each extension phase 1 leaves
open is materialized as an order-\(12\) tournament \(L + S\) and decided
by the same engine, again at unit margin, where a witness settles it
positively. About one extension in \(200{,}000\) reached this phase, and
every one of them is \(5\)-inducible with unit margin.

\textbf{Phase 3, plain majority.} Only a \emph{majority} refutation
would be a counterexample, so any instance negative at unit margin is
re-run without the margin restriction. Phase 3 was never invoked: every
tournament was settled at unit margin, so the weaker question never had
to be asked.

\subsubsection{C.3 One tournament of each converse
pair}\label{c.3-one-tournament-of-each-converse-pair}

\(T\) is \(5\)-inducible if and only if its converse \(\bar{T}\) is
(Section 1). Moreover
\(\overline{L + S} = \bar{L} + (V(L) \setminus S)\), so settling all
\(2{,}048\) extensions of \(L\) settles all \(2{,}048\) of \(\bar{L}\)
as well. We therefore keep one tournament of each converse pair, by the
rule that \(L\) is kept if and only if
\(\mathrm{canon}(L) \le \mathrm{canon}(\bar{L})\). This keeps exactly
one of each pair together with every self-converse tournament,
\((D_{11} + S_{11})/2\) in all, where \(S_{11}\) is the number of
self-converse order-\(11\) classes; this saves a factor of
\(\lessapprox 2\). We also check the number of tournaments kept is
\((D_{11} + S_{11})/2\).

Total runtime: \(2{,}615.6\) core-hours. The search code, the per-part
records and the roll-up are at \texttt{cluster/jz\_n12cover} in the
repository of Section 5.

\subsection{Appendix D. Exact enumeration
counts}\label{appendix-d.-exact-enumeration-counts}

\setcounter{table}{0}\renewcommand{\thetable}{D.\arabic{table}}

In this table, \(D_n\), \(R_n\) and \(S_n\) are the numbers of
isomorphism classes of all, of regular, and of self-converse tournaments
on \(n\) vertices. They match OEIS A000568 (OEIS Foundation Inc.~2026a),
A096368 (OEIS Foundation Inc.~2026c), and A002785 (OEIS Foundation
Inc.~2026b).

\begin{longtable}[]{@{}rrrr@{}}
\caption{Numbers of isomorphism classes of all, regular and
self-converse tournaments.}\tabularnewline*
\toprule\noalign{}
\(n\) & \(D_n\) & \(R_n\) & \(S_n\) \\*
\midrule\noalign{}
\endfirsthead
\toprule\noalign{}
\(n\) & \(D_n\) & \(R_n\) & \(S_n\) \\*
\midrule\noalign{}
\endhead
\bottomrule\noalign{}
\endlastfoot
\(11\) & \(903{,}753{,}248\) & \(1{,}223\) & \(279{,}968\) \\*
\(12\) & \(154{,}108{,}311{,}168\) & --- & \(1{,}492{,}288\) \\*
\(13\) & --- & \(1{,}495{,}297\) & \(95{,}458{,}560\) \\*
\(15\) & --- & \(18{,}400{,}989{,}629\) & --- \\*
\end{longtable}

\subsection{\texorpdfstring{Appendix E. Locally transitive tournaments,
at unit margin, for every odd
\(k \ge 3\)}{Appendix E. Locally transitive tournaments, at unit margin, for every odd k \textbackslash ge 3}}\label{appendix-e.-locally-transitive-tournaments-at-unit-margin-for-every-odd-k-ge-3}

\setcounter{table}{0}\renewcommand{\thetable}{E.\arabic{table}}

\setcounter{figure}{0}\renewcommand{\thefigure}{E.\arabic{figure}}

A tournament is \emph{locally transitive} if for every vertex \(v\) both
\(N^{+}(v)\) and \(N^{-}(v)\) induce transitive subtournaments.
Equivalently it is \emph{round}: the vertices admit a cyclic order
\(v_0, \dots, v_{n-1}\) in which every \(N^{-}(v) \cup \{v\}\) and
\(\{v\} \cup N^{+}(v)\) is an interval, so that, orienting the cyclic
order along the arcs, each vertex beats exactly the
\(d_i = |N^{+}(v_i)|\) vertices that follow it:
\(N^{+}(v_i) = \{v_{i+1}, \dots, v_{i+d_i}\}\), indices modulo \(n\).
The equivalence is due to Brouwer (1980), who derived the cyclic order
on the way to enumerating the class, and to Alspach and Tabib (1982),
who state it for tournaments under the name \emph{domination
orientable}.

\textbf{Theorem E.1.} \emph{Every locally transitive tournament is
\(3\)-inducible with unit margin, and hence \(k\)-inducible with unit
margin for every odd \(k \ge 3\).}

\begin{figure}[!htbp]
\centering
\includegraphics[width=\linewidth]{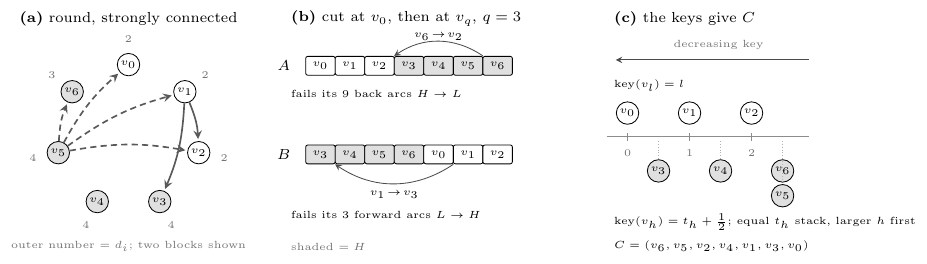}
\caption{The construction of Appendix E on a round tournament with out-degree sequence
$(2, 2, 2, 4, 4, 4, 3)$, for which $q = 3$. \textbf{(a)} The cyclic order, each $v_i$ labelled with its
out-degree $d_i$, so that its out-neighbours are $v_{i+1}, \dots, v_{i+d_i}$; the two blocks drawn
are those of $v_1 \in L$ and $v_5 \in H$, and that of $v_5$ wraps, with back-targets $v_0, v_1, v_2$.
\textbf{(b)} $A$ and $B$ are the same cyclic order cut at $v_0$ and at $v_q$, each failing exactly the
arcs that cross its own cut, $9$ back arcs and $3$ forward arcs; one of each is drawn.
\textbf{(c)} The keys, $C$ being the vertices in decreasing order of key.
A vertex of $L$ sits at its index and one of $H$ half a unit past its ceiling, so
reading leftwards puts $v_h$ before $v_l$ exactly when $v_h \to v_l$; $v_5$ and $v_6$ share a
ceiling and the tie-break separates them, giving $C = (v_6, v_5, v_2, v_4, v_1, v_3, v_0)$.}
\end{figure}

A locally transitive tournament that is not strongly connected is
transitive: a strongly connected component \(C\) with \(|C| \ge 3\)
contains a directed triangle, and any vertex outside \(C\) either beats
all of \(C\) or loses to all of it, contradicting local transitivity. A
transitive tournament is \(1\)-inducible at margin \(1\), its single
voter supporting every arc. Adding an order together with its reverse
adds one supporter and one opponent to each arc, raising every support
by \(1\) and \(k\) by \(2\) and so leaving every margin \(2c(e) - k\)
unchanged. Padding with the identity order \((1, 2, \dots, n)\) and its
reverse therefore gives unit margin at \(k = 3\), and repeating gives
every odd \(k\).

Let \(T\) be strongly connected and round, and let \(d_i\) be the
out-degree of vertex \(i\); Figure E.1 illustrates what follows, on
seven vertices. Cutting the cyclic order at \(v_0\) gives the linear
order \(A = (v_0, \dots, v_{n-1})\). We write \(r_i = i + d_i\), an
integer deliberately \emph{not} reduced modulo \(n\). For each \(v_i\),
the out-neighbours are then \(v_{i+1}, \dots, v_{r_i}\) with indices
reduced only when they are read as vertices. The sequence \(r\) is
non-decreasing, since if \(r_{i+1} < r_i\) then \(v_{r_{i+1}+1}\) is an
out-neighbour of \(v_i\), while its own interval includes \(v_i\), a
contradiction. Let \(H = \{i : r_i \ge n\}\) be the vertices with an
out-neighbour that the cut defined by \(A\) placed \emph{before} them.
As \(r\) is non-decreasing, \(H = \{q, \dots, n-1\}\) for some \(q\);
write \(L = \{0, \dots, q-1\}\). Neither set is empty: \(H = \emptyset\)
would make \(v_{n-1}\) a sink and \(L = \emptyset\) would force
\(d_0 \ge n\), so neither occurs. For \(h \in H\) put \(t_h = r_h - n\),
so that the out-neighbours of \(v_h\) placed before it by the cut are
exactly \(v_0, \dots, v_{t_h}\): we call them the \emph{back-targets} of
\(v_h\), and \(t_h\) its \emph{ceiling}. By monotonicity of \(r\),
\(t_q \le \dots \le t_{n-1}\). Every back-target lies in \(L\), that is
\(t_{n-1} < q\): otherwise \(v_{n-1} \to v_q\), while \(r_q \ge n\)
implies \(v_q \to v_{n-1}\). The arcs between \(L\) and \(H\) therefore
divide into the \emph{back arcs} going from \(H\) to \(L\), and the
\emph{forward arcs} going from \(L\) to \(H\); and for \(h \in H\) and
\(l \in L\) we have \(v_h \to v_l\) exactly when \(l \le t_h\).

We give each vertex a \emph{key}: \(\mathrm{key}(v_l) = l\) for
\(l \in L\) and \(\mathrm{key}(v_h) = t_h + \tfrac12\) for \(h \in H\),
ties among equal ceilings broken in favour of the larger index, and let
\(C\) list the vertices in decreasing order of key. Integer and
half-integer keys never collide, so that tie-break is the only one
needed, the keys are pairwise distinct and \(C\) is a linear order; and
for \(h \in H\) and \(l \in L\),
\(\mathrm{key}(v_h) > \mathrm{key}(v_l) \iff t_h + \tfrac12 > l \iff l \le t_h \iff v_h \to v_l\),
so \(C\) places \(v_h\) before \(v_l\) exactly when \(v_h \to v_l\). We
now take the three orders \(A = (v_0, \dots, v_{n-1})\),
\(B = (v_q, \dots, v_{n-1}, v_0, \dots, v_{q-1})\), which is the same
cyclic order cut at \(v_q\) instead of at \(v_0\), and \(C\). Every arc
is now easily seen to have support exactly two. \emph{Arcs with both
endpoints inside \(L\) or inside \(H\)} are only supported by \(A\) and
\(B\); \emph{forward arcs \(v_l \to v_h\)} are only supported by \(A\)
and \(C\); and \emph{back arcs \(v_h \to v_l\)} are only supported by
\(B\) and \(C\), so the profile realizes \(T\) at unit margin. The
padding argument above extends this margin to any odd \(k \ge 3\).
\(\blacksquare\)

\subsection{\texorpdfstring{Appendix F. Criticality at \(21\)
vertices}{Appendix F. Criticality at 21 vertices}}\label{appendix-f.-criticality-at-21-vertices}

\setcounter{table}{0}\renewcommand{\thetable}{F.\arabic{table}}

This appendix carries the detail behind Section 3.6. Four of the \(110\)
vertex-transitive tournaments on \(21\) vertices are unit-margin
obstructions (Section 3.7). We name them by their automorphism groups,
all of order \(21\): \(C_a\) and \(C_b\) have the cyclic group
\(\mathbb{Z}_{21}\), and \(F_a\) and \(F_b\) the nonabelian group
\(\mathbb{Z}_7 \rtimes \mathbb{Z}_3\).

\textbf{Vertex-critical but maximally not arc-critical.} Arc-criticality
implies vertex-criticality, and the converse fails here as widely as it
can. Each of the four is \emph{vertex-critical} at unit margin, and for
\(C_a\) every single-arc reversal remains a unit-margin obstruction.

\textbf{Arc-criticality is a property of the arc, not just the
tournament.} All four have ten arc orbits of size \(21\); we sweep one
representative per orbit at unit margin and get results that differ by
tournament \emph{and} by arc, \texttt{U} marking a reversal that is
still not unit-margin inducible and \texttt{S} one that is. \(C_b\)
splits five orbits to five, so exactly half its arcs are critical and we
call it \emph{arc-semi-critical}; with \(P_{23}\) arc-critical
unrestricted and \(C_a\) critical nowhere, all three possibilities occur
among our tournaments. Every one of the \(15\) reversals that remains an
obstruction belongs to a tournament with \emph{cyclic automorphism
group}, while all \(20\) orbits of the two with the nonabelian group of
order \(21\) are inducible.

\begin{longtable}[]{@{}
  >{\raggedright\arraybackslash}p{(\linewidth - 6\tabcolsep) * \real{0.2500}}
  >{\raggedright\arraybackslash}p{(\linewidth - 6\tabcolsep) * \real{0.2500}}
  >{\raggedright\arraybackslash}p{(\linewidth - 6\tabcolsep) * \real{0.2500}}
  >{\raggedright\arraybackslash}p{(\linewidth - 6\tabcolsep) * \real{0.2500}}@{}}
\caption{Single-arc reversals of the four vertex-transitive unit-margin
obstructions on \(21\) vertices, one per arc orbit.}\tabularnewline*
\toprule\noalign{}
\begin{minipage}[b]{\linewidth}\raggedright
tournament
\end{minipage} & \begin{minipage}[b]{\linewidth}\raggedright
\(\mathrm{Aut}\)
\end{minipage} & \begin{minipage}[b]{\linewidth}\raggedright
the ten orbits
\end{minipage} & \begin{minipage}[b]{\linewidth}\raggedright
obstructions
\end{minipage} \\*
\midrule\noalign{}
\endfirsthead
\toprule\noalign{}
\begin{minipage}[b]{\linewidth}\raggedright
tournament
\end{minipage} & \begin{minipage}[b]{\linewidth}\raggedright
\(\mathrm{Aut}\)
\end{minipage} & \begin{minipage}[b]{\linewidth}\raggedright
the ten orbits
\end{minipage} & \begin{minipage}[b]{\linewidth}\raggedright
obstructions
\end{minipage} \\*
\midrule\noalign{}
\endhead
\bottomrule\noalign{}
\endlastfoot
\(C_a\) & \(\mathbb{Z}_{21}\) & \texttt{U\ U\ U\ U\ U\ U\ U\ U\ U\ U} &
\(10\) \\*
\(C_b\) & \(\mathbb{Z}_{21}\) & \texttt{U\ S\ U\ S\ S\ U\ U\ S\ S\ U} &
\(5\) \\*
\(F_a\), \(F_b\) & \(\mathbb{Z}_7 \rtimes \mathbb{Z}_3\) &
\texttt{S\ S\ S\ S\ S\ S\ S\ S\ S\ S} & \(0\) \\*
\end{longtable}

\subsection{Appendix G. The search below 22
vertices}\label{appendix-g.-the-search-below-22-vertices}

\setcounter{table}{0}\renewcommand{\thetable}{G.\arabic{table}}

This appendix summarizes the search for an obstruction on fewer than
\(22\) vertices, which would improve the upper bound of Section 3.2. It
has found none.

The search runs downward, because \(5\)-inducibility is inherited by
subtournaments: restricting a witness for \(T\) to a subset of the
vertices gives a witness for the subtournament. An obstruction on fewer
vertices therefore lies below some larger one, and the natural place to
look is among the vertex deletions of known obstructions. A host is
settled positively as soon as one base state yields a witness, which is
cheap, and negatively only by a complete search, which is not; so each
level is first screened for witnesses, and only hosts that resist the
screen are searched further.

The starting points are the \(36\) doubly regular tournaments on \(23\)
vertices other than \(P_{23}\), all of which are obstructions (Section
3.7). \(P_{23}\) itself is excluded, since \(P_{23} - v\) is
\(5\)-inducible (Section 3.6) and so, by inheritance, is every
tournament below it. Up to isomorphism and converse, the levels below
the \(36\) are as follows.

\begin{longtable}[]{@{}
  >{\raggedleft\arraybackslash}p{(\linewidth - 6\tabcolsep) * \real{0.0533}}
  >{\raggedleft\arraybackslash}p{(\linewidth - 6\tabcolsep) * \real{0.0533}}
  >{\raggedleft\arraybackslash}p{(\linewidth - 6\tabcolsep) * \real{0.0533}}
  >{\raggedright\arraybackslash}p{(\linewidth - 6\tabcolsep) * \real{0.8400}}@{}}
\caption{The levels below the \(36\) doubly regular tournaments on
\(23\) vertices, up to isomorphism and converse.}\tabularnewline*
\toprule\noalign{}
\begin{minipage}[b]{\linewidth}\raggedleft
vertices
\end{minipage} & \begin{minipage}[b]{\linewidth}\raggedleft
deleted
\end{minipage} & \begin{minipage}[b]{\linewidth}\raggedleft
classes
\end{minipage} & \begin{minipage}[b]{\linewidth}\raggedright
status
\end{minipage} \\*
\midrule\noalign{}
\endfirsthead
\toprule\noalign{}
\begin{minipage}[b]{\linewidth}\raggedleft
vertices
\end{minipage} & \begin{minipage}[b]{\linewidth}\raggedleft
deleted
\end{minipage} & \begin{minipage}[b]{\linewidth}\raggedleft
classes
\end{minipage} & \begin{minipage}[b]{\linewidth}\raggedright
status
\end{minipage} \\*
\midrule\noalign{}
\endhead
\bottomrule\noalign{}
\endlastfoot
\(22\) & \(1\) & \(378\) & \(8\) searched completely, all obstructions
(Section 3.2) \\*
\(21\) & \(2\) & \(4{,}108\) & \(1{,}192\) settled, all
\(5\)-inducible \\*
\(20\) & \(3\) & \(28{,}724\) & all \(5\)-inducible \\*
\end{longtable}

At \(22\) vertices all \(378\) classes were first screened at up to
\(50\) base states each, without a single witness, and the eight
complete searches were chosen from among them. At \(21\) vertices the
\(1{,}192\) settled classes include all \(506\) two-vertex deletions of
\(\Delta_6\) and \(\Delta_{28}\), so every \(21\)-vertex subtournament
of either parent is \(5\)-inducible, and in particular
\(\Delta_6 - v_{5}\) and \(\Delta_{28} - v_{5}\) are vertex-critical.
The other settled classes are a stratified random sample of \(400\), all
\(53\) classes below \(\Delta_{11}\), and every class obtained by
deleting \(v_5\) together with one other vertex, the deletion that
produced those two obstructions, and \(19\) classes settled by inserting
the missing vertex into a known witness of one of their \(20\)-vertex
subtournaments, as in phase 1 of Appendix C.2; the remaining \(2{,}916\)
classes are open. At \(20\) vertices every class is \(5\)-inducible, and
so, by inheritance, is every tournament on \(19\) vertices or fewer
below the \(36\).

Two obstructions on \(25\) vertices from other families were descended
the same way: \(P_{27}\) with two vertices deleted, and the circulant
tournament on \(25\) vertices in which every arc lies in six directed
triangles. Neither is \(5\)-inducible, by complete searches, and every
\(21\)-vertex subtournament of either is \(5\)-inducible, so neither
family yields an obstruction below \(22\).

\textbf{Other families on \(21\) vertices.} We also searched families on
\(21\) vertices that a symmetry or a construction makes enumerable, all
of them close to doubly regular. A regular tournament on \(21\) vertices
cannot be doubly regular, since \((21 - 3)/4\) is not an integer; we
call it \emph{near-doubly regular} if every pair of vertices has \(4\)
or \(5\) common out-neighbours. The table below lists every family we
searched on \(21\) vertices, again up to isomorphism and converse, with
the number of classes settled; the vertex-transitive tournaments on
\(21\) vertices, a family enumerated in full, are covered in Section
3.7. Every settled class is \(5\)-inducible, by a witness verified
independently of the search; none was settled negatively.

\begin{longtable}[]{@{}
  >{\raggedright\arraybackslash}p{(\linewidth - 4\tabcolsep) * \real{0.8272}}
  >{\raggedleft\arraybackslash}p{(\linewidth - 4\tabcolsep) * \real{0.0864}}
  >{\raggedleft\arraybackslash}p{(\linewidth - 4\tabcolsep) * \real{0.0864}}@{}}
\caption{The families searched on \(21\) vertices, up to isomorphism and
converse.}\tabularnewline*
\toprule\noalign{}
\begin{minipage}[b]{\linewidth}\raggedright
family on \(21\) vertices
\end{minipage} & \begin{minipage}[b]{\linewidth}\raggedleft
classes
\end{minipage} & \begin{minipage}[b]{\linewidth}\raggedleft
settled
\end{minipage} \\*
\midrule\noalign{}
\endfirsthead
\toprule\noalign{}
\begin{minipage}[b]{\linewidth}\raggedright
family on \(21\) vertices
\end{minipage} & \begin{minipage}[b]{\linewidth}\raggedleft
classes
\end{minipage} & \begin{minipage}[b]{\linewidth}\raggedleft
settled
\end{minipage} \\*
\midrule\noalign{}
\endhead
\bottomrule\noalign{}
\endlastfoot
two vertices deleted from a doubly regular tournament on \(23\) vertices
& \(4{,}108\) & \(1{,}192\) \\*
four vertices deleted from the circulant tournament on \(25\) vertices &
\(286\) & \(286\) \\*
six vertices deleted from \(P_{27}\) & \(150\) & \(150\) \\*
near-doubly regular, rigid, drawn at random & \(78\) & \(78\) \\*
near-doubly regular, with a bisector and an automorphism of order \(5\)
& \(28\) & \(28\) \\*
near-doubly regular, with a fixed-point-free automorphism of order \(3\)
& \(31\) & \(31\) \\*
\end{longtable}

The \(31\) classes of the last family are likely, but not confirmed, to
be all of it: the enumeration found no new class after its first \(225\)
solutions, and we stopped it at \(4{,}567\) without running it to
completion. Two of the \(31\) are bisectable and the other \(29\) are
not. In the two families with an automorphism, several classes are not
\(5\)-inducible at unit margin and need a majority witness: \(12\) of
the \(28\) with a bisector, and \(30\) of the \(31\) with an
automorphism of order \(3\).

\subsection{Appendix H. The obstructions on 22
vertices}\label{appendix-h.-the-obstructions-on-22-vertices}

\setcounter{table}{0}\renewcommand{\thetable}{H.\arabic{table}}

This appendix gives the obstructions of Sections 3.2 and 3.3 so that
each can be checked without the catalogue of McKay (2026). Each is given
in the catalogue's own format, a single string of \(231\) bits giving
the upper triangle of its adjacency matrix row by row: for every
\(u < w\), in the labelling of the parent with the deleted vertex
removed, \(1\) if the \(u\)-th vertex beats the \(w\)-th and \(0\)
otherwise. Each string is printed in three lines of \(77\), below the
order of the automorphism group and, when it is non-trivial, a minimal
set of generators in cycle notation, in the same labels.

\noindent

\begin{minipage}{\linewidth}
\textbf{$\Delta_1 - v_{1}$}\quad {\small $|\mathrm{Aut}| = 5$, generated by (2 9 11 7 5)(3 8 4 12 6)(13 22 15 16 20)(14 17 19 23 18)}\\[2pt]
{\small\ttfamily 11111000001111110000001100011110001111100100111000010110011111100110110010101\\
11011010101100110101011100101100110100110000101110010101101101010101100011100\\
10101011001011101101001100010110101011010101111000001101010001000101011110011}
\end{minipage}
\medskip

\noindent

\begin{minipage}{\linewidth}
\textbf{$\Delta_1 - v_{2}$}\quad {\small $|\mathrm{Aut}| = 5$, generated by (1 19 12 23 8)(3 6 4 18 14)(5 16 15 13 7)(9 22 11 20 21)}\\[2pt]
{\small\ttfamily 11111111110000000000001100011110001111100100111000010110011111100110110010101\\
11011010101100110101011100101100110100110000101110010101101101010101100011100\\
10101011001011101101001100010110101011010101111000001101010001000101011110011}
\end{minipage}
\medskip

\noindent

\begin{minipage}{\linewidth}
\textbf{$\Delta_1 - v_{10}$}\quad {\small $|\mathrm{Aut}| = 55$, generated by (2 9 11 7 5)(3 8 4 12 6)(13 22 15 16 20)(14 17 19 23 18) and\\
(1 18 4 14 23 19 12 6 8 17 3)(2 16 21 7 13 5 11 9 22 15 20)}\\[2pt]
{\small\ttfamily 11111111110000000000011111000011111100000011000111000111110010011000010110011\\
11110110110010101110101010110011010111100101100110001100001011101010110110100\\
10101011001011101101001100010110101011010101111000001101010001000101011110011}
\end{minipage}
\medskip

\noindent

\begin{minipage}{\linewidth}
\textbf{$\Delta_{11} - v_{2}$}\quad {\small $|\mathrm{Aut}| = 5$, generated by (1 21 12 22 9)(3 14 13 7 16)(4 17 18 15 6)(8 11 10 23 19)}\\[2pt]
{\small\ttfamily 11111111110000000000001110110001100001111001010110011101001110101100010111100\\
11101011001010011111011010111011000101011100111000101011001011010110001110101\\
11001001011100011101010001011110010011001010101000101010110010001011101110111}
\end{minipage}
\medskip

\noindent

\begin{minipage}{\linewidth}
\textbf{$\Delta_{11} - v_{5}$}\quad {\small $|\mathrm{Aut}| = 5$, generated by (1 21 12 22 9)(3 14 13 7 16)(4 17 18 15 6)(8 11 10 23 19)}\\[2pt]
{\small\ttfamily 11111111110000000000011110000011111100000011011000110000111101010110011101001\\
11101011001010011111011010111011000101011100111000101011001011010110001110101\\
11001001011100011101010001011110010011001010101000101010110010001011101110111}
\end{minipage}
\medskip

\noindent

\begin{minipage}{\linewidth}
\textbf{$\Delta_{11} - v_{20}$}\quad {\small $|\mathrm{Aut}| = 5$, generated by (1 21 12 22 9)(3 14 13 7 16)(4 17 18 15 6)(8 11 10 23 19)}\\[2pt]
{\small\ttfamily 11111111111000000000011111000001111110000011101100011000011100101011001110101\\
11010110001011100111010110010101111101101011101000101011100110001010110011101\\
01100011010111001000111000111101000101110010010010101100010010110100101101111}
\end{minipage}
\medskip

\noindent

\begin{minipage}{\linewidth}
\textbf{$\Delta_6 - v_{5}$}\quad {\small $|\mathrm{Aut}| = 1$}\\[2pt]
{\small\ttfamily 11111111110000000000011110000011111100000011001101001001111001100010000111101\\
11011100010110110111111110111100001010110010010111010101010111010011010101101\\
10110011001101001101011000110011011010101000111001001010100110010011100101110}
\end{minipage}
\medskip

\noindent

\begin{minipage}{\linewidth}
\textbf{$\Delta_{28} - v_{5}$}\quad {\small $|\mathrm{Aut}| = 1$}\\[2pt]
{\small\ttfamily 11111111110000000000011110000011111100000011101001000101011101010100001011101\\
11100110010110111011111110111100000110110100011100011100010101110101010110011\\
01101010110001101101011000110101011011010000111000100110000011010011100101110}
\end{minipage}
\medskip

\noindent

\begin{minipage}{\linewidth}
\textbf{$\Delta_1 - v_{10}$ with the arc between $v_2$ and $v_3$ reversed}\quad {\small $|\mathrm{Aut}| = 1$}\\[2pt]
{\small\ttfamily 11111111110000000000001111000011111100000011000111000111110010011000010110011\\
11110110110010101110101010110011010111100101100110001100001011101010110110100\\
10101011001011101101001100010110101011010101111000001101010001000101011110011}
\end{minipage}
\medskip

\end{document}